\documentclass[a4paper,12pt]{article}

\newlength{\lead}
\usepackage{leading}
\leading{\lead}

\usepackage{calc}
\newlength{\abz}
\newlength{\measure}
\usepackage[a4paper, textwidth=1.15\measure, 
bindingoffset=18pt,
bottom=2.5cm,
headsep=2\lead, footskip=2\lead]{geometry}
\usepackage[utf8]{inputenc}
\usepackage[english]{babel}
\usepackage{graphicx}
\usepackage[titletoc,title]{appendix}
\usepackage{setspace} 
\usepackage[nolist,printonlyused,smaller]{acronym} 

\usepackage[babel]{csquotes}

\usepackage{amsmath}
\usepackage{amssymb}
\usepackage{amsfonts}
\usepackage{mathrsfs} 
\usepackage{mathtools} 
\usepackage{bm} 
\usepackage{dsfont} 
\usepackage{hyperref}
\usepackage{tikz}
\usetikzlibrary{arrows,decorations.pathreplacing}
\usepackage{empheq} 
\usepackage{extarrows}
\usepackage[figuresright]{rotating}
\usepackage{slashed}

\usepackage{geometry} 

\numberwithin{equation}{section}

\newcommand{\ita}[1]{\textit{#1}}
\newcommand{\mrm}[1]{\mathrm{#1}}
\newcommand{\lef}{\left}
\newcommand{\ri}{\right}
\newcommand{\e}{\mathrm{e}}

\newcommand{\foh}{\frac{1}{2}}
\newcommand{\Tr}{\operatorname{Tr}}

\newcommand{\D}{\mathrm{d}}

\newcommand{\sgo}{\sqrt{g}}

\newcommand{\sgbo}{\sqrt{\bar{g}}}
\newcommand{\sgb}[1]{\sqrt{\bar{g}(#1)}}

\newcommand{\intD}{\int\!\mathcal{D}}
\newcommand{\mB}{\mathscr{B}}
\newcommand{\mD}{\mathcal{D}}
\newcommand{\mF}{\mathscr{F}}
\newcommand{\mE}{\mathscr{E}}

\newcommand{\mT}{\mathscr{T}}

\newcommand{\mM}{\mathscr{M}}

\newcommand{\mK}{\mathscr{K}}
\newcommand{\dd}{\D^d}

\newcommand{\what}[1]{\widehat{#1}}

\newcommand{\bg}{\begin{equation}}
\newcommand{\eg}{\end{equation}}
\newcommand{\bgo}{\begin{equation*}}
\newcommand{\ego}{\end{equation*}}

\newcommand{\spl}[1]{\begin{split}#1\end{split}} 

\def\m{_{\mu}}
\def\n{_{\nu}}
\def\M{^{\mu}}
\def\N{^{\nu}}
\def\a{_{\alpha}}

\def\A{^{\alpha}}
\def\B{^{\beta}}
\def\r{_{\rho}}
\def\s{_{\sigma}}

\def\mn{_{\mu\nu}}
\def\MN{^{\mu\nu}}
\def\ab{_{\alpha\beta}}
\def\AB{^{\alpha\beta}}
\def\rs{_{\rho\sigma}}

\def\RS{^{\rho\sigma}}

\def\MA{^{\mu\alpha}}

\def\NB{^{\nu\beta}}

\makeatletter
\newcommand*\bigcdot{\mathpalette\bigcdot@{.5}}
\newcommand*\bigcdot@[2]{\mathbin{\vcenter{\hbox{\scalebox{#2}{$\m@th#1\bullet$}}}}}
\makeatother

\DeclarePairedDelimiter\bra{\langle}{\rvert} 
\DeclarePairedDelimiter\ket{\lvert}{\rangle} 
\DeclarePairedDelimiterX\braket[2]{\langle}{\rangle}{#1 \delimsize\vert #2} 

\newcommand{\gl}[1]{\eqref{eq:#1}}
\newcommand{\Gl}[1]{Eq.~\eqref{eq:#1}}

\begin{document}

\begin{acronym}


\acro{qft}[QFT]{quantum field theory}

\acro{rhs}[RHS]{right-hand side}

\acro{lhs}[LHS]{left-hand side}

\acro{rs}[RS]{rigid spacetime}

\acro{sc}[SC]{selfconsistent}




\acro{uv}[UV]{ultraviolet}

\acro{ir}[IR]{infrared}

\acro{brst}[BRST]{Becchi-Rouet-Stora-Tyutin}




\end{acronym}

\newgeometry{ 
a4paper, 
total={170mm,257mm}, 
left=20mm, 
top=00mm, 
}

\begin{titlepage}

\title{
\begin{flushright}
{\footnotesize MITP-21-039}
\end{flushright}
\vspace*{3.5cm}
Background Independent Field Quantization with\\
\mbox{}\\
Sequences of Gravity-Coupled Approximants II:\\
\mbox{}\\
Metric Fluctuations\\
\mbox{}
}

\date{}

\author{Maximilian Becker\footnote{E-mail: \texttt{bemaximi@uni-mainz.de}}$\ $ and Martin Reuter\footnote{E-mail: \texttt{reutma00@uni-mainz.de}}\\[3mm]
{\small Institute of Physics, 
Johannes Gutenberg University Mainz,}\\[-0.2em]
{\small Staudingerweg 7, D--55099 Mainz, Germany}
}

\maketitle
\thispagestyle{empty}

\vspace{2mm}
\begin{abstract}

\noindent
We apply the new quantization scheme outlined in Phys. Rev. D102 (2020) 125001 to explore the influence which quantum vacuum fluctuations of the spacetime metric exert on the universes of Quantum Einstein Gravity, which is regarded an effective theory here. The scheme promotes the principle of Background Independence to the level of the regularized precursors of a quantum field theory (``approximants'') and severely constrains admissible regularization schemes. Without any tuning of parameters, we find that the zero point oscillations of linear gravitons on maximally symmetric spacetimes do not create the commonly expected cosmological constant problem of a cutoff-size curvature. On the contrary, metric fluctuations are found to reduce positive curvatures to arbitrarily tiny and ultimately vanishing values when the cutoff is lifted. This suggests that flat space could be the distinguished groundstate of pure quantum gravity. Our results contradict traditional beliefs founded upon background-dependent calculations whose validity must be called into question therefore.

\end{abstract}

\setstretch{1.2} 

\end{titlepage}

\restoregeometry

\newpage


\section{Introduction}

Uniting the laws of gravity with the principles of quantum mechanics is commonly believed to present a major challenge that is beset with formidable difficulties and obstacles. Among the culprits that contributed perhaps most to this belief are certain notorious divergences; they are taken responsible for disasters such as the perturbative non-renormalizability of quantum General Relativity, or the excedingly wrong theoretical expectations concerning the observed cosmological constant, to mention just two.\\ \indent
On a more positive note, divergences do not always deserve their bad reputation. Sometimes it happens that, if one carefully listens to the physics message they tell us, all of a sudden we see that actually the divergences are not the \textit{cause} of the problem, but rather the key to its \textit{solution}. The way of how Asymptotic Safety overcomes the renormalizability problem can serve as an example of such a metamorphosis~\cite{wein,Percacci:2017fkn,Reuter:2019byg}.\\ \indent
This paper is devoted to a similar phenomenon connected to the cosmological constant problem~\cite{Weinberg:1988cp}. More precisely, we are interested in the way quantum vacuum fluctuations (``zero point oscillations'') influence the geometry of spacetime. Considering quantized metric gravity (without matter here), the main role will be played by the purportedly fatal quartic divergences which are at the heart of the grossly false estimates for the expected induced cosmological constant.\\ \indent 
As it will turn out, if we take them seriously as the sign of a strong, physically real quantum effect, then they can indeed be converted from a problem to a solution. Making essential use of the lessons implied by Background Independence~\cite{Ashtekar:2014kba,Giu-BI,Reuter:2019byg} we shall demonstrate that those divergences, rather than giving rise to a spacetime curvature that is \ita{too large} by many orders of magnitude, actually do their best to \textit{flatten} spacetime. The mechanism we are going to describe here might help then to explain why the universe we live in is so much larger than its ``natural'' scale, the Planck length.\\

\noindent
\textbf{(1) Approximant systems.} The present investigation extends work initiated in the context of classical gravity~\cite{N1} towards the realm of quantum gravity. In the companion paper~\cite{N1}, henceforth denoted by~[I], we outlined a novel scheme for the quantization of quantum matter fields coupled to classical gravity. We proposed that quantum field theories should be regularized by \textit{sequences of quasi-physical systems.} Those systems, called ``approximants,'' are required to possess a well defined number of degrees of freedom, $\mathbf f$ say, which represent a finite subset of the excitations that can be carried by the field. Typically, they are chosen as $\mathbf f$ of its harmonic normal modes.\\ \indent 
While the quantization of such finite approximants, denoted by ``$\mathsf{App}(\mathbf f)$,'' is straightforward, in principle, we insisted that all of them are coupled to gravity; in particular the $\mathbf f$ quantum degrees of freedom must be able to dynamically backreact on the metric of the classical spacetime they live in. Each one of the systems $\mathsf{App}(\mathbf f)$ determines its preferred background geometry self-consistently as the solution of a certain tadpole condition, typically a kind of semiclassical Einstein equation.\\ \indent\newpage
The total configuration of an approximant $\mathsf{App}(\mathbf f)$ is described by both a quantum mechanical state $\Psi_{\mathbf f}$ of the quantized matter degrees of freedom, and a classical metric \textit{which is self-consistent with respect to this particular quantum state}. Symbolically, we write
\bg
\mathsf{App}(\mathbf f)\sim\Psi_{\mathbf f}\otimes\text{metric}(\Psi_{\mathbf f}) \, .
\eg
The construction of a continuum limit would then amount to designing an infinite sequence of such systems $\{\mathsf{App}(\mathbf f)\}$ which possesses a meaningful and physically interesting limit:
\bg
\mathsf{App}(\mathbf f)\xrightarrow{\mathbf f\to\infty}\Psi_\mrm{QFT}^\text{\acs{sc}}\otimes\bar g\mn^\text{\acs{sc}}(\Psi_\mrm{QFT}) \, .
\label{eq:1.10}
\eg
The important point here is that the limit \gl{1.10} produces not only a state $\Psi^\text{\acs{sc}}_\mrm{QFT}$ of the now fully quantized matter field theory, but in addition also a dynamically determined background metric $\bar g\mn^\text{\acs{sc}}$ which is self-consistent precisely when the matter system is in the state $\Psi^\text{\acs{sc}}_\mrm{QFT}$ of the \ac{qft}.\\

\noindent
\textbf{(2) Requirements on an improved quantization scheme.} The coupling to gravity and their individual dynamical backreaction on the spacetime metric is a defining property of the approximant systems. This property may be regarded an extension of the general principle of Background Independence~\cite{Ashtekar:2014kba} to the \textit{regularized precursor} of the quantum field theory. In fact, in~[I] we advocated the following three general requirements concerning the quantization scheme:
\begin{center}
\begin{minipage}{0.4\textwidth}
\begin{itemize}
\item[\textbf{(R1)}] Background Independence
\item[\textbf{(R2)}] Gravity-coupled approximants
\item[\textbf{(R3)}] $N$-type cutoffs \, .
\end{itemize}
\end{minipage}
\end{center}
\enlargethispage{\baselineskip}
The motivation for the last requirement, \textbf{(R3)}, is to strictly disentangle the concepts of, respectively, \textit{regularization}, i.e., the construction of sequences of approximants, and of \textit{scale setting}. The defining property of \textit{$N$-type cutoffs} is that they must not employ the metric in defining or selecting the approximants' degrees of freedom. As a consequence, the numerical value associated with a specific $N$-cutoff is a dimensionless number, and therefore it avoids defining any particular length or mass scale.\\ \indent
In~[I] we discussed in detail why the combination of \textbf{(R2)} and \textbf{(R3)} allows us to take limits of $\{\mathsf{App}(\mathbf f)\}$ sequences that could not be considered within the standard schemes. Intuitively speaking, the idea behind the \textit{quasi-physical} systems $\mathsf{App}(\mathbf f)$ is the expectation that the closer the approximants come to being quantum mechanical systems that can actually be realized in nature, the more likely it is that they approach a meaningful limit $\lim_{\mathbf f\to\infty}\mathsf{App}(\mathbf f)$. In this sense, \textbf{(R2)} expresses that a system can be ``physical'' only if it lives in harmony with gravity, i.e., if it inhabits a spacetime geometry which is determined dynamically by the energy and momentum of the inhabitants. Additionally, \textbf{(R3)} allows us to employ the concept of a spacetime metric only \textit{within} the individual systems $\mathsf{App}(\textbf f)$, $\mathbf f$ fixed, \textit{but not for the very definition of the approximants}. In the standard regularization schemes with dimensionful cutoffs those two roles of the metric often get intermingled, something we consider unacceptable when the metric is dynamical itself.\\

\noindent
\textbf{(3) $N$-type cutoffs.} Using an $N$-cutoff, the regularization of a generic functional integral
\bg
\label{eq:1.10-2}
Z=\int_\mF\!\mD(\chi)\,\e^{-S[\chi]}
\eg
over all fields $\chi$ in a given (Hilbert) space of functions $\mF$ proceeds as follows. In $\mF$, we pick a basis $\mB=\{w\a\ |\ \alpha\in I\}$, i.e., $\mF=\mrm{span}\,\mB$ so that every field has an expansion $\chi(x)=\sum_{\alpha\in\ I}c\a w\a(x)$ in terms of basis functions $w\a$ labeled by (a list of) indices from a certain index set $I$. Regarding the measure, we interpret the ill-defined formal product $\mD(\chi)\text{``$=$''}\prod_{x\in M}\D\chi(x)$ over all points of the (fixed, differentiable) spacetime manifold $M$, as $\prod_{\alpha\in I}\D c\a$, and then we install the ``$N$-cutoff'' in the resulting combination of infinitely many ordinary integrals.\\ \indent
To do so, we introduce a one-parameter family of subsets $\mB_N\subset\mB$ labeled by a dimensionless number $N\in\mathds{N}$.\footnote{With minor modifications, the case $N\in[0,\infty)$ is possible as well, see~[I].} The subsets are required to satisfy
\bg
\label{eq:1.15}
\mB_0=\emptyset\quad,\quad\mB_\infty=\mB\quad\text{and}\quad N_2>N_1\Rightarrow\mB_{N_2}\supset\mB_{N_1} \, .
\eg
Crucially, the specification of admissible sequences of sub-bases,
\bg
\label{eq:1.16}
\mB_0\subset\mB_1\subset\mB_2\subset\cdots\subset\mB_N\subset\cdots\subset\mB
\eg
must not involve any data other than the index set $I$. Every $\mB_N$ must be described as
\bg
\label{eq:1.17}
\mB_N=\{w\a\ |\  \alpha\in I_N\}
\eg
with $I_N\subset I$ a suitable subset of index values. Letting $\mF_N=\mrm{span}\,\mB_N$ yields a sequence of finite dimensional function spaces $\mF_N\subset\mF$, and a corresponding sequence of regularized integrals
\bg
\label{eq:1.18}
Z_N=\int_{\mF_N}\!\mD(\chi)\,e^{-S[\chi]} \, .
\eg
They extend over progressively larger subspaces $\mF_N$ when the cutoff parameter $N=0,1,2,\cdots,\infty$ is increased.\\ \indent
We consider \gl{1.18} as the defining partition function (generating functional) of the approximant system $\mathsf{App}(N)\equiv\mathsf{App}(\mathbf f(N))$. Its degrees of freedoms are constituted by a selection of $\mathbf f\equiv\mathbf f(N)$ modes of the field $\chi$. The number $\mathbf f$ is given by the dimensionality of $\mF_N$ and increases monotonically with $N=0,1,2,\cdots,\infty$.\\ \indent
The deeper reason behind the requirement of an $N$-type cutoff is to ensure that the individual approximants really come close to viable physical systems in their own right, at least to the extent that they possess a clearcut number of degrees of freedom. In many standard schemes, in particular those using a dimensionful cutoff scale, this would not be the case. Typically such schemes employ a metric in defining the regularized functional integral, and as we explained in~[I] already, it is usually impossible then to interpret this integral as the partition function of some quantum mechanical system. The reason of this failure is that when gravity is dynamical, changes of the metric  can reshuffle matter degrees of freedom back and forth between regularizations with different values of the cutoff scale. Therefore, if we try to approximate an infinite dimensional quantum system (a \acs{qft}) by a sequence of smaller physical systems, we must avoid standard regularization schemes of this sort.\\ \indent
Considerable care is required when the basis elements $w\a$ are taken to be eigenfunctions of some kinetic operator $\mK$ which depends on the background metric:
\bg
\label{eq:1.20}
\mK[\bar g]\, w\a[\bar g](x)=\lambda\a[\bar g]\,w\a[\bar g](x)\quad,\quad\alpha\in I \, .
\eg
Even in this case, where the eigenvalues $\lambda\a[\bar g]$ and the eigenfunctions $w\a[\bar g]$ are manifestly $\bar g$-dependent, it is still possible to define a bona fide $N$-cutoff in terms of
\bg
\label{eq:1.21}
\mB_N[\bar g]=\{w\a[\bar g]\ |\ \alpha\in I_N\}
\eg
and $\mF_N[\bar g]=\mrm{span}\,\mB_N[\bar g]$. Here, the key point is that, while the basis functions as such are metric dependent, \ita{the index set enumerating them, $I_N$, is not}. We refer to~[I] for a detailed discussion of this important class of $N$-cutoffs which is somewhat subtle to deal with.\\ 

\noindent
\textbf{(4) Application: vacuum fluctuations and the cosmological constant.} In~[I] we applied the new quantization scheme based upon the three requirements \textbf{(R1)}, \textbf{(R2)} and \textbf{(R3)} to a concrete physical question, namely the effect of quantum vacuum fluctuations on the curvature of spacetime. Considering a free scalar matter field on a classical Euclidean spacetime $(S^d, \bar g\mn)$, we constructed infinite sequences of appropriate quantum mechanical systems, $\{\mathsf{App}(\mathbf f)\ |\ \mathbf f=\mathbf f(N)\ ,\ N=0,1,2,\cdots,\infty\}$, and for each approximant separately we determined its respective self-consistent metric $(\bar g_{\mathbf f}^\text{\acs{sc}})\mn$ from the total state $\mathsf{App}(\mathbf f)\sim\Psi_{\mathbf f}^\text{\acs{sc}}\otimes(\bar g_{\mathbf f}^\text{\acs{sc}})\mn$. The latter is dictated by a coupled system of equations consisting of a $\bar g\mn$-dependent Schr\"odinger equation for $\Psi_{\mathbf f}$, and a generalized Einstein equation with a quantum stress tensor $T\mn[\Psi_{\mathbf f}]$ for the metric.\\ 

\begin{figure}[ht]
	\centering
  \includegraphics[width=0.9\textwidth]{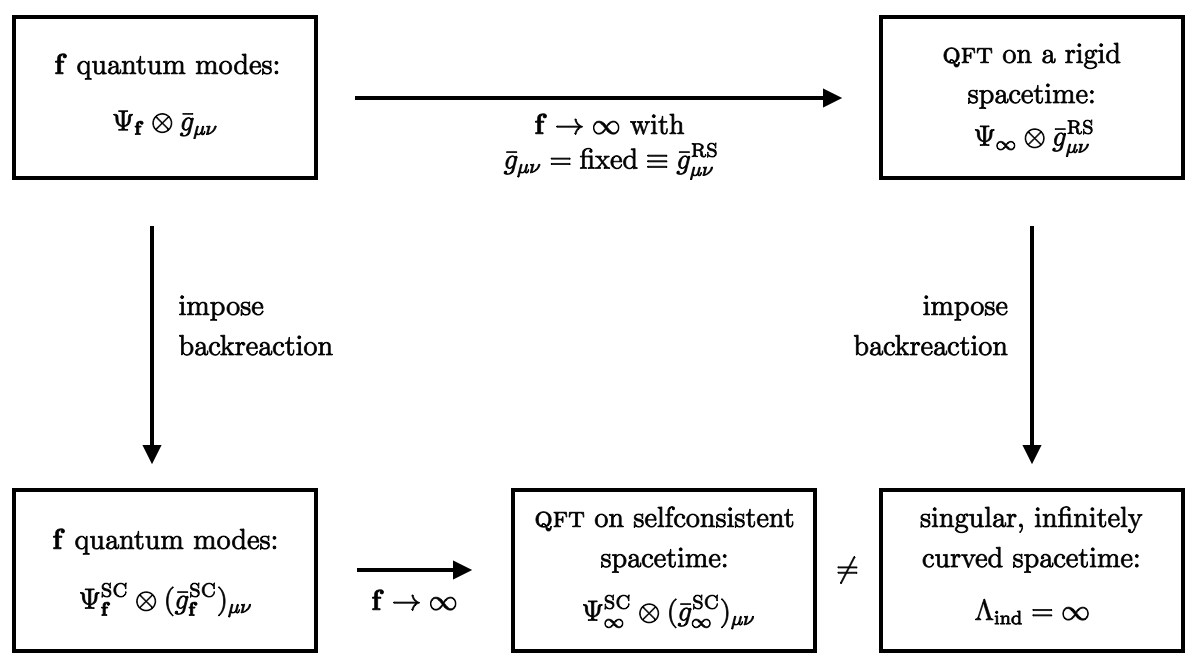}
	\caption{The diagram illustrates that the inclusion of the gravitational backreaction (vertical arrows) does not commute with the limit $\mathbf f\to\infty$ (horizontal arrows). The upper left box represents a regularized matter theory on a spacetime that carries an arbitrary metric unrelated to the matter systems. The traditional quantization procedure follows the path ``first right, then down.'' Thereby, the continuum limit $\mathbf f\to\infty$ is performed on a fixed, rigid spacetime with metric $\bar g^\text{\acs{rs}}\mn$. This leads to an infinite cosmological constant, and when the backreaction is included, to a singular spacetime with infinite curvature. The new quantization scheme, the path ``first down, then right,'' instead results in a well-defined quantum state $\Psi^\text{\acs{sc}}_\infty$ on a nonsingular spacetime. It arises automatically equipped with a flat metric $(\bar g_\infty^\text{\acs{sc}})\mn$. (Taken from [I].)}
	\label{fig:D}
\end{figure}

Investigating the continuum limit \gl{1.10} of sequences for which $\lim_{\mathbf f\to\infty}\Psi_{\mathbf f}^\text{\acs{sc}}=\Psi_\mrm{QFT}^\text{\acs{sc}}$ is the no-particle state revealed a true surprise then: Contrary to general belief, adding further field modes to the approximant does not increase, but rather decreases the curvature of spacetime, and in the continuum limit $\lim_{\mathbf f\to\infty}\mathsf{App}(\mathbf f)$ the approximants' spacetimes converge to \ita{flat space} actually.\\ \indent
These results are strikingly different from those of the well-known standard calculations which, following Pauli~\cite{Pauli-Calc}, sum up zero-point energies and try to take the \acs{qft} limit \ita{before} the gravitational backreaction on the matter system is taken into account.\footnote{Reviews of the cosmological constant problem and of earlier attempts at its solution include Refs. \cite{Weinberg:1988cp,Pauli-Calc,Carroll:2000fy,Straumann:1999ia,Straumann:2002tv,pad-CC}. Concerning more recent foundational work on conceptual aspects of the problem, unrelated to our approach however, we refer the reader to \cite{Hollands:2004xv,Wang:2017oiy,Sudarsky,Carlip:2018zsk}.} The latter procedure gives rise to an (infinitely) large induced cosmological constant $\Lambda_\mrm{ind}=\infty$, and a correspondingly large (divergent) curvature of spacetime. Indeed, our main conclusion in ref.~[I] was that 
\ita{the crucial steps of taking the continuum limit $\mathbf f\to\infty$, and including gravity's backreaction on matter do not commute.}\\ \indent
Thus it appears that the notorious ``cosmological constant problem,'' at least to the extent it relates to zero point oscillations, is an unphysical artefact originating from a wrong way of taking the continuum limit. According to the new scheme based upon \textbf{(R1,2,3)}, the \acs{qft} limit $\mathbf f\to\infty$ is taken \ita{after} the inclusion of the backreaction. It describes a well-defined vacuum state of the matter system that inhabits a non-singular spacetime. The latter turns out completely flat without any tuning of parameters or further ado.\\ \indent
In Figure \ref{fig:D}, this situation is summarized schematically. For further details we must refer to~[I].

\noindent
\textbf{(5) Quantum fluctuations of the metric.} While already in~[I] gravity was treated dynamical, but at the classical level only, the purpose of the present paper is to generalize the analysis by a quantum mechanical treatment also of the gravitational field itself. More precisely, we are going to quantize the quadratic metric fluctuations implied by the Einstein-Hilbert action, and we investigate how their vacuum oscillations affect the self-consistent spacetimes of the corresponding approximants.\\ \indent
Technically, we shall perform a linear background split of the metric, $g\mn=\bar g\mn+h\mn$, whereby $\bar g\mn$ is a dynamically adjustable background metric, as in~[I], while the fluctuation field $h\mn$ plays a role similar to that of a matter field. We describe the approximants $\mathsf{App}(\mathbf f)$ by suitable regularized action functionals of the form $\Gamma[h\mn,\chi;\bar g\mn]$ with $\chi\equiv(\chi_a)$ an arbitrary collection of additional (matter) fields. Backgrounds $\bar g\mn\equiv\bar g\mn^\text{\acs{sc}}$ are \ita{self-consistent} if they allow the effective field equations implied by $\Gamma$ to admit the solution $h\mn=0$. Finding those special background geometries requires solving the tadpole condition
\bg
\label{eq:1.30}
\frac{\delta}{\delta h\mn}\Gamma[h,\chi;\bar g]\Bigg|_{h=0,\,\bar g=\bar g^\text{\acs{sc}}}=0
\eg
together with the matter field equations $\delta\Gamma/\delta\chi_a=0$.\\ \indent
In this paper we are going to focus on the case where no genuine matter fields are present, so that the set $(\chi_a)$ comprises only the Faddeev-Popov ghosts that arise by gauge fixing the general coordinate invariance. We shall analyze whether the spacetimes predicted by \ita{pure Quantum Einstein Gravity} are plagued by a cosmological constant problem due to the zero point oscillations of the geometry.\\ 

\noindent
\textbf{(6) This paper.}The remaining sections of this paper are organized as follows. In Section 2 we prepare the stage by introducing the classical graviton and ghost system that we are going to consider, and we set up the saddle point approximation of the functional integral over all metrics.\\ \indent 
In Section 3 we start the construction of the approximants, and focus in particular on the (integrated trace oft the) effective stress tensor by means of which they backreact on gravity; a number of additional technical developments are necessary here to make the approximants autonomous physical systems.\\ \indent 
The construction of the approximants is completed in Section 4, where we specialize the setting for spherical (Euclidean) spacetimes and give a detailed description of the truncated bases that underlie the degrees of freedom governed by $\mathsf{App}(N)$.\\ \indent 
Finally, in Section 5 our main results are derived and interpreted; there we install the self-consistent gravitational backreaction in all approximants, and analyze the resulting series $\{\mathsf{App}(N)\}$, being particularly interested in the field theory limit $N\to\infty$. Section 6 contains a brief summary of our results.\\ \indent
Various details on the tensor harmonics of $S^d$, and on certain spectral sum representations are relegated to three appendices.\\ \indent
In the sequel we shall frequently refer to the companion paper~[I] in which further details can be found.

\section{Vacuum fluctuations of the geometry}

In this preparatory section we assume that we are given some generic, $d$-dimensional Riemannian manifold $(M, \bar g\mn)$ and we set up the classical and quantum theory of metric fluctuations $h\mn$ relative to the classical background metric $\bar g\mn$. Performing a linear split between background and fluctuation field, we represent the total metric as $g\mn=\bar g\mn+h\mn$. We think of $h\mn$ as a kind of matter field, while $\bar g\mn$ plays exactly the same r\^ole as in~[I], reviewed  in the Introduction.

\subsection{Classical gravitons and ghosts}

As for its classical dynamics, we assume that $g\mn$, and hence $h\mn$, are governed by the Einstein-Hilbert action
\bg
\label{eq:2.11}
S_\mrm{EH}[g]=2\kappa^2\!\!\int\!\dd x\,\sgo\lef(-R+2\Lambda_\mrm{b}\ri)
\eg
with
\bg
\label{eq:2.12}
\kappa^2\equiv\frac{1}{32\pi G}
\eg
and an arbitrary bare cosmological constant, $\Lambda_\mrm{b}$.\\

\noindent
\textbf{(1) Gauge fixing.} In order to deal with the diffeomorphism invariance of $S_\mrm{EH}$ we promote it to a \ac{brst} gauge-fixed ``quantum action'' in the standard way~\cite{Nak-Oj}. Introducing diffeomorphism ghosts $C\M$ and antighosts $\bar C\m$, respectively, it reads
\bg
\label{eq:2.13}
S[h,\bar C,C;\bar g]=S_\mrm{EH}[\bar g+h]+S_\mrm{gf}[h;\bar g]+S_\mrm{gh}[h,\bar C,C;\bar g]
\eg
where $S_\mrm{gf}$ and $S_\mrm{gh}$ denote the gauge fixing and ghost terms, respectively. We adopt the formalism without a Nakanishi-Lautrup auxiliary field and employ a gauge fixing action which is quadratic in $h\mn$,
\bg
\label{eq:2.14}
S_\mrm{gf}[h;\bar g]=\kappa^2\!\!\int\!\dd x\,\sgbo\,\bar g\MN\!\lef(\mathcal{F}\m\AB[\bar g]h\ab\ri)\!\lef(\mathcal{F}\n^{\gamma\delta}[\bar g]h_{\gamma\delta}\ri)\, ,
\eg
and involves the differential operator
\bg
\label{eq:2.15}
\mathcal{F}\m\AB[\bar g]\equiv\delta\m\B\,\bar g^{\alpha\gamma}\bar D_\gamma-\foh\,\bar g\AB\bar D\m \, .
\eg
The concomitant ghost action has the structure
\bg
\label{eq:2.16}
S_\mrm{gh}[h,\bar C,C;\bar g]=-\sqrt{2}\!\int\!\dd x\,\sgbo\,\bar C\m{\mM[\bar g+h,\bar g]\M}\n C\N \, ,
\eg
where the corresponding Faddeev-Popov operator reads explicitly
\bg
\label{eq:2.17}
{\mM[g,\bar g]\M}\n=\bar g^{\mu\rho}\bar g^{\sigma\lambda}\,\bar D_{\lambda}\lef(g_{\rho\nu} D\s+g_{\sigma\nu}D\r\ri)-\bar g^{\rho\sigma}\bar g^{\mu\lambda}\,\bar D_\lambda g_{\sigma\nu} D\r \, .
\eg
Here, $D\m$ and $\bar D\m$ denote the covariant derivatives built from $g\mn$ and $\bar g\mn$, respectively. It can be verified that the total quantum action \gl{2.13} enjoys an on-shell \ac{brst} invariance~\cite{reuter}.\\

\noindent
\textbf{(2) Matter-like fields.} In this paper we often regard the dynamical fields $(h\mn,\bar C\m,C\M)$ as a special sort of ``matter'' which inhabits the classical spacetime $(M,\bar g\mn)$. Separating off the purely gravitational part, the corresponding ``matter action'' is given by
\bg
\label{eq:2.18}
\spl{
S_\mrm{M}[h,\bar C,C;\bar g]&\equiv S[h,\bar C,C;\bar g]-S_\mrm{EH}[\bar g]\\
&=S_\mrm{EH}[\bar g+h]-S_\mrm{EH}[\bar g]+S_\mrm{gf}[h;\bar g]+S_\mrm{gh}[h,\bar C,C;\bar g] \, .
}
\eg
Here, we encounter an action functional of the bi-metric type at the classical level already. It depends on two independent metrics, $\bar g\mn$ and $\bar g\mn+h\mn\equiv g\mn$, respectively.\\

\noindent
\textbf{(3) Classical tadpole condition.} For later comparison it is instructive to write down the self-consistency condition \gl{1.30} for the classical case, i.e., when $\Gamma$ is given by the bare action $S$ from \Gl{2.13}. We have to solve
\bg
\label{eq:2.19}
\frac{\delta}{\delta h\mn}S[h,\bar C,C;\bar g]\Bigg|_{h=0,\,\bar g=\bar g^\text{\acs{sc}}}=0
\eg
along with two further equations in which $C\M$ and $\bar C\m$ are varied. The simple $\bar C(\cdots)C$-structure of the ghost action entails that the latter equations are solved by $C\M=0=\bar C\m$. Combining this solution with the first equation, \gl{2.19}, the latter becomes
\bg
\label{eq:2.20}
\spl{
0&=\frac{\delta}{\delta h\mn}\Big(S_\mrm{EH}[\bar g+h]+S_\mrm{gf}[h;\bar g]\Big)\Bigg|_{h=0,\,\bar g=\bar g^\text{\acs{sc}}}\\
&=\frac{\delta}{\delta h\mn}S_\mrm{EH}[\bar g+h]\Bigg|_{h=0,\,\bar g=\bar g^\text{\acs{sc}}} \, .
}
\eg
In the second line, we exploited that $S_\mrm{gf}$ is bilinear in $h\mn$ and supplies no contribution to the functional derivative at $h=0$ therefore. If we now reinstate the variable $g\mn=\bar g\mn+h\mn$, the condition \gl{2.20} turns into the expected form, namely
\bg
\label{eq:2.21}
\frac{\delta}{\delta g\mn}S_\mrm{EH}[g]\Bigg|_{g=g^\text{\acs{sc}}}=0 \, .
\eg \indent
To summarize, classically the quasi-matter fields $(h,\bar C,C)$ governed by the full fledged \ac{brst} invariant action live in a self-consistent background geometry if the ghost fields vanish, and the metric $\bar g\mn$ is a stationary point of $S_\mrm{EH}$, i.e., of the classical action functional before its promotion to a \ac{brst} invariant action.\\

\noindent
\textbf{(4) Quadratic actions: free gravitons and ghosts.} Let us expand the ``matter'' action $S_\mrm{M}$ given in \Gl{2.18} around the point $(h=0,\bar C=0=C)$ up to terms of second order in the dynamical fields $(h,\bar C,C)$. The result has the structure
\bg
\label{eq:2.22}
S_\mrm{M}[h,\bar C,C;\bar g]=S_\mrm{FG}[h;\bar g]+S_{\mrm{F}\mrm{gh}}[\bar C,C;\bar g]+(\text{linear})+O(3) \, .
\eg
Here, ``O(3)'' stands for terms of cubic and higher orders, while the ``linear'' terms vanish precisely when the background is a self-consistent one. Furthermore, $S_\mrm{FG}$ and $S_{\mrm{F}\mrm{gh}}$ denote the ``Free Graviton'' and ``Free ghost'' actions, which are bilinear in the respective dynamical fields:
\bg
\label{eq:2.23}
S_\mrm{FG}[h;\bar g]\equiv\foh\!\int\!\dd x\,\sgbo\ h\MN{\mK[\bar g]\mn}\RS h\rs \, ,
\eg
\bg
\label{eq:2.23-1}
\spl{
S_{\mrm{F}\mrm{gh}}[\bar C,C;\bar g]&\equiv S_\mrm{gh}[0,\bar C,C;\bar g]\\
&=-\sqrt{2}\!\int\!\dd x\,\sgbo\ \bar C\m{\mM[\bar g,\bar g]\M}\n C\N \, .
}
\eg
The graviton's non-minimal kinetic operator $\mK\equiv\mK_0+\mK_1$ comprises the two contributions $\mK_0$ and $\mK_1$ which stem from the terms in the action $\propto\sgo R$ and $\propto\sgo$, respectively. They read, for a generic background metric,
\bg
\label{eq:2.24}
\spl{
\mK_0[\bar g]\rs{}\MN&=\frac{1}{16\pi G}\lef[-\bar D^2 K\rs{}\MN+V\rs{}\MN\ri]\\
\mK_1[\bar g]\rs{}\MN&=\frac{1}{16\pi G}\lef[-2\Lambda_\mrm{b}\, K\rs{}\MN\ri]
}
\eg
with 
\bg
\label{eq:2.25}
K\rs{}\MN\equiv\frac{1}{4}\lef[\delta\r\M\delta\s\N+\delta\s\M\delta\r\N-\bar g\MN\bar g\rs\ri] \, .
\eg
The potential-type curvature terms, built from $\bar g\mn$, are given by
\bg
\label{eq:2.26}
\spl{
V\rs{}\MN\equiv\bar R\, K\rs{}\MN&+\foh\lef[\bar g\MN\bar R\rs+\bar g\rs\bar R\MN\ri]\\
&-\frac{1}{4}\lef[\delta\r\M\bar R\N{}\s+\delta\M\s\bar R\N{}\r+\delta\r\N\bar R\M{}\s+\delta\N\s\bar R\M{}\r\ri]-\foh\lef[\bar R\N{}\r{}\M{}\s+\bar R\N{}\s{}\M{}\r\ri] \, .
}
\eg
Here and in the following indices are always raised and lowered with the background metric $\bar g\mn$. We emphasize in particular that the fundamental dynamical variable is the tensor field $h\mn$ with two lower indices. Thus equations such as \gl{2.23} should be interpreted as a short hand for
\bg
\label{eq:2.27}
S_\mrm{FG}[h;\bar g]=\foh\!\int\!\dd x\,\sgbo\ h\ab\bar g^{\alpha\mu}\bar g^{\beta\nu}\mK[\bar g]\mn{}\RS h\rs
\eg
with $h\ab\bar g^{\alpha\mu}\bar g^{\beta\mu}$ abbreviated by $h\MN$.

\subsection{Background Independent quantum theory}

Turning to the quantum theory now, we would like to compute arbitrary correlation functions  $\langle\what{\varphi}(x_1)\cdots\what{\varphi}(x_n)\rangle_{\bar g}$ where $\what{\varphi}\equiv(\what{h}\mn,\what{\bar C}\m,\what{C}\M)$ denotes the operatorial counterpart of $(h\mn,\bar C\m,C\M)\equiv\varphi$. The nontrivial challenge is to compute those $n$-point functions \ita{as a functional of the background metric $\bar g\mn$}, or phrased in a different way, to quantize the fluctuations on all possible geometries \ita{in one go}.\\ \indent
In this subsection we briefly review the corresponding formal, i.e., un-regularized path integrals representing the relevant generating functionals. Later on we outline how to regularize them by means of an $N$-cutoff.\\ \indent
Since the actual meaning is clear from the context, we use the notation $\what{\varphi}\equiv(\what{h}\mn,\what{\bar C}\m,\what{C}\M)$ both for operators and integration variables. Thus, after coupling the dynamical fields $\what{\varphi}\equiv(\what{h}\mn,\what{\bar C}\m,\what{C}\M)$ to sources $(t\MN,\sigma\M,\bar\sigma\m)$, we can write down the following (formal) functional integral for the generating functional of their connected correlators:
\bg
\label{eq:2.30}
\spl{
\exp\Big(W[t,\sigma,\bar\sigma;\bar g]\Big)=\intD(\what{h};\bar g)&\mD(\what{C};\bar g)\mD(\what{\bar C};\bar g)\exp\Bigg(-S[\what{h},\what{\bar C},\what{C};\bar g]\\
&+\int\!\dd x\,\sgbo\,\lef\{t\MN\what{h}\mn+\bar\sigma\m\what{C}\M+\sigma\M\what{\bar C}\m\ri\}\Bigg) \, .
}
\eg
This functional integral involves the \ac{brst} invariant action \gl{2.13}. It can be seen as the result of formally applying the Faddeev-Popov trick to $\int\!\mD\what{g}\,\e^{-S_\mrm{EH}[\what{g}]}$, thereby using a gauge fixing condition of the background field type~\cite{back-method}. Furthermore, in \Gl{2.20} we interpret $\mD(\what{h};\bar g)\mD(\what{C};\bar g)\mD(\what{\bar C};\bar g)$ as the \ac{brst}-invariant measure found by Fujikawa~\cite{grav-measure}:
\bg
\label{eq:2.31}
\mD(\what{h};\bar g)\mD(\what{C};\bar g)\mD(\what{\bar C};\bar g)=\prod_x\bar g(x)^{\frac{(d-4)(d+1)}{8}-\frac{d}{2}}\lef(\prod_{\mu\geq\nu}\D\what{h}\mn(x)\ri)\!\lef(\prod_\alpha\D\what{C}\A(x)\D\what{\bar C}\a(x)\ri) \, .
\eg
Note its dependence on the background metric via the chracteristic power of the determinant $\bar g(x)\equiv\det(\bar g\ab(x))$.\\ \indent 
The formal construction of the effective action for the ``matter'' fields on the $(M,\bar g)$ background is completed by performing a Legendre-Fenchel transformation with respect to the sources, at fixed $\bar g\mn$. This leads us to the generating functional for the 1PI Green's functions of $\what{h}\mn$ and the ghosts:
\bg
\label{eq:2.35}
\Gamma[h\mn,\bar C\m,C\M;\bar g\mn]\equiv\Gamma[g\mn,\bar g\mn,\bar C\m,C\M] \, .
\eg
Here, the field expectation values are denoted by 
\bg
\label{eq:2.36}
h\mn\equiv\langle\what{h}\mn\rangle_{\bar g}\quad,\quad\bar C\m\equiv\langle\what{\bar C}\m\rangle_{\bar g}\quad\text{and}\quad C\M\equiv\langle\what{C}\M\rangle_{\bar g} \, ,
\eg
and in the case of the total metric $\what{g}\mn=\bar g\mn+\what{h}\mn$ we write $g\mn\equiv\langle\what{g}\mn\rangle=\bar g\mn+h\mn$.\\ \indent 
The desired $n$-point functions can be obtained by repeated differentiation of \Gl{2.35},
\bg
\label{eq:2.37}
\langle\what{\varphi}(x_1)\cdots\what{\varphi}(x_n)\rangle_{\bar g}\propto\lef(\frac{\delta}{\delta\varphi}\ri)^n\Gamma[\varphi;\bar g] \, .
\eg
As we stressed already, we are particularly interested in their $\bar g$-dependence. The one-point function of the metric fluctuation, $\langle\what{h}\mn(x)\rangle_{\bar g}$, is of special significance in this respect. Its vanishing is the defining property of a self-consistent background geometry in the quantum mechanical case. Since $\langle\what{g}\mn\rangle_{\bar g}=\bar g\mn+\langle\what{h}\mn\rangle_{\bar g}$ it implies that the expectation value of the metric operator equals exactly the prescribed background metric if the latter is a self-consistent one:
\bg
\label{eq:2.38}
\langle\what{g}\mn\rangle_{\bar g}=\bar g\mn\quad\Leftrightarrow\quad\langle\what{h}\mn\rangle_{\bar g}=0\quad\quad\quad\text{if}\quad\bar g=\bar g^\text{\acs{sc}} \, .
\eg
This requirement is known as a tadpole condition. In terms of $\Gamma$, it consists of the equation
\bg
\label{eq:2.39}
\frac{\delta}{\delta h\mn(x)}\Gamma[h,\bar C,C;\bar g]\Bigg|_{h=0,\, \bar g=\bar g^\text{\acs{sc}}}=0 \, ,
\eg
and, also imposing $\langle\what{C}\M\rangle_{\bar g}=0=\langle\what{\bar C}\m\rangle_{\bar g}$, two similar equations that involve ghost derivatives. The tadpole condition \gl{2.39} is the quantum mechanical analogue of the classical equation \gl{2.19} discussed above. Replacing the bare action $S$ by the functional $\Gamma$, the condition of self-consistency carries over from the classical to the quantum world in a simple and natural way.

\section{One-loop action and functional measure}

The loop-wise expansion of integrals like \gl{2.30} is well known. In this section we recall the saddle point approximation of the path integral about the point of vanishing fluctuation fields, and then highlight the importance of the functional measure in this connection. We elaborate on a number of technical points which, while unimportant in typical calculations on flat spacetime, become nontrivial in a Background Independent setting. In particular it is essential then to carefully keep track of all occurrences of $\bar g\mn$ in the functional integral.

\subsection{Tadpole condition at one loop}

At one-loop order, and for vanishing ghost arguments, the effective action has the structure
\bg
\label{eq:2.50}
\Gamma[h,0,0;\bar g]=S_\mrm{EH}[\bar g+h]+\Gamma_\mrm{1L}[\bar g+h]+O(\text{2 loops})
\eg
with
\bg
\label{eq:2.50-1}
\Gamma_\mrm{1L}[\bar g]=\Gamma_\mrm{FG}[\bar g]+\Gamma_{\mrm{F}\mrm{gh}}[\bar g]
\eg
whereby the free graviton ($\mrm{FG}$) and free ghost ($\mrm{F}\mrm{gh}$) effective actions $\Gamma_\mrm{FG}$ and $\Gamma_{\mrm{F}\mrm{gh}}$, respectively, are given by two Gaussian integrals which we write down in a moment.\\ \indent
Before, let us return for a moment to the effective field equation, the tadpole condition that governs the backreaction of the quantum vacuum fluctuations on their habitat, the classical spacetime $(M,\bar g)$, and that decides about whether or not a certain background geometry is self-consistent.\\ \indent
The effective field equations to be considered consist of \Gl{2.39} coupled to the corresponding equations for the ghosts. Solving the latter by choosing $\bar C\m=0=C\M$, it remains to solve
\bg
\label{eq:drei.1}
\frac{\delta}{\delta h\mn(x)}\Gamma[h,0,0;\bar g]\Bigg|_{h=0,\, \bar g=\bar g^\text{\acs{sc}}}=0 \, .
\eg
Henceforth we shall employ the one-loop approximation \gl{2.50} for the effective action. As a result, \Gl{drei.1} assumes the form of a classical-looking Einstein equation,
\bg
\label{eq:3.20}
\bar R\MN-\foh\bar g\MN\bar R+\Lambda_\mrm{b}\,\bar g\MN=(8\pi G)\,T\MN[\bar g]
\eg
whereby the stress tensor
\bg
\label{eq:3.40}
T\MN[\bar g](x)\equiv-\frac{2}{\sgbo}\frac{\delta}{\delta\bar g\mn(x)}\Gamma_\mrm{1L}[\bar g]
\eg
originates from the one-loop terms $\Gamma_\mrm{1L}=\Gamma_\mrm{FG}+\Gamma_\mrm{Fgh}$.

\subsection{Functional measure and Gaussian integrals}

The saddle point approximation provides us with the following Gaussian functional integrals over gravitons and ghosts whose only interaction is with the background gravitational field:\footnote{In cases where the positioning of indices is critical we indicate the intended index structure of the respective geometric quantities by dots.}
\bg
\label{eq:2.50-2}
\e^{-\Gamma_\mrm{FG}[\bar g]}=\intD(\what{h}_{\bigcdot\,\bigcdot};\bar g_{\bigcdot\,\bigcdot})\,\e^{-S_\mrm{FG}[\what{h}_{\bigcdot\,\bigcdot};\bar g_{\bigcdot\,\bigcdot}]}
=\intD_1\!\!\lef[\bar g^{(d-4)/(4d)}\what{h}_{\bigcdot\,\bigcdot}\ri]\,\e^{-S_\mrm{FG}[\what{h}_{\bigcdot\,\bigcdot};\bar g_{\bigcdot\,\bigcdot}]} \, ,
\eg
\bg
\label{eq:2.50-3}
\spl{
\e^{-\Gamma_\mrm{Fgh}[\bar g]}&=\intD(\what{C}\,{}^{\bigcdot};\bar g_{\bigcdot\,\bigcdot})\mD(\what{\bar C}_{\bigcdot};\bar g_{\bigcdot\,\bigcdot})\,\e^{-S_\mrm{Fgh}[\what{\bar C}_{\bigcdot},\what{C}\,{}^{\bigcdot};\bar g_{\bigcdot\,\bigcdot}]}\\
&=\intD_1\lef[\bar g^{(d+2)/(4d)}\what{C}\,{}^{\bigcdot}\ri]\!\mD_1\!\!\lef[\bar g^{(d-2)/(4d)}\what{\bar C}_{\bigcdot}\ri]\e^{-S_\mrm{Fgh}[\what{\bar C}_{\bigcdot},\what{C}\,{}^{\bigcdot};\bar g_{\bigcdot\,\bigcdot}]} \, .
}
\eg
Here, we factorized the measure \gl{2.31} and expressed it in terms of the na\"ive measure $\mD_1$. By definition, the latter does not involve any factors of the determinant $\bar g(x)=\det(\bar g_{\bigcdot\,\bigcdot}(x))$:
\bg
\label{eq:2.50-4}
\spl{
\mD_1[h_{\bigcdot\,\bigcdot}]&=\prod_x\prod_{\mu\geq\nu}\D h\mn(x)\\
\mD_1[C\,{}^{\bigcdot}]&=\prod_x\prod\m\D C\M(x)\\
\mD_1[\bar C_{\bigcdot}]&=\prod_x\prod\m\D\bar C\m(x) \, .
}
\eg
Here and in the following we omit the carets over the integration variables.\\ \indent
In order to make the above measures and integrals well-defined we restrict them to a finite number of spacetime points and attach integration variables like $h\mn(x)$ only to the sites of some lattice or triangulation. The precise details of this intermediate regularization do not matter as it anyhow will be replaced by the $N$-cutoff at a later stage.\\ \indent
The actions appearing under the above integrals, $S_\mrm{FG}$ and $S_\mrm{Fgh}$, are given by the bilinear functionals \gl{2.23} and \gl{2.23-1}, respectively, which involve the operators $\mK[\bar g]$ and $\mM[\bar g,\bar g]$.\\ \indent 
Let us stress at this point that the approximation which leads to \gl{2.50-2} neglects interactions among gravitons obviously. The nonrenormalizability of quantum General Relativity is not an issue here, and a status as an effective theory is sufficient.\\ \indent
\enlargethispage{\baselineskip}
In the rest of this subsection we are now going to evaluate the integrals explicitly. Because of the non-standard measures they are not (yet) strictly Gaussian. A certain amount of additional care is required here, not only because of the field-dependent measures, but also in view of our special needs for what concerns the definition of approximants respecting \textbf{(R1,2,3)}; this will be the topic of the next subsection then.\\

\noindent
\textbf{(1) The spectral problem.} Background Independence forces us to evaluate the generating functional and hence $\Gamma_\mrm{FG}[\bar g]$, say, for arbitrary background metrics $\bar g\mn$. Therefore, the evaluation of \gl{2.50-2} starts out from the assumption that the spectral problem of the operator $\mK[\bar g]$ has been solved for arbitrary $\bar g\mn$,
\bg
\label{eq:eig-1}
\mK[\bar g]\mn{}\AB\, u_{nm}(x)\ab=\mF_n[\bar g]\,u_{nm}(x)\ab \, ,
\eg
so that we know the metric-dependent eigenfunctions and eigenvalues. As in~[I] we assume a compact spacetime, hence a discrete spectrum, and for every $\bar g$ we enumerate the eigenvalues in increasing order: $\mF_0\leq\mF_1\leq\mF_2\leq\mF_3\leq\cdots$. The eigenfunctions $u_{nm}$ form a complete set of symmetric tensor fields on $M$, satisfying the orthogonality and completeness relations
\bg
\label{eq:eig-2}
\int\!\dd x\,\sgbo\,\bar g^{\mu\alpha}(x)\bar g^{\nu\beta}(x)\, u^*_{nm}(x)\ab\ u_{\bar n\bar m}(x)\mn =\delta_{n\bar n}\delta_{m\bar m} \, ,
\eg
\bg
\label{eq:eig-3}
\sum_{n,m}\bar g^{\alpha\rho}\bar g^{\beta\sigma}\,u_{nm}(x)\rs\,u^*_{nm}(y)\mn=\frac{\delta(x-y)}{\sgb{x}}\delta\A_{(\mu}\delta\B_{\nu)} \, .
\eg
The symmetrization in \gl{eig-3} is defined to include a factor of $1/2$.\\

\noindent
\textbf{(2) Densitized fields.} As a first step towards actually computing the functional integrals we introduce the densitized fluctuation fields
\bg
\label{eq:2.50-10}
\spl{
f\mn&\equiv\bar g^{(d-4)/(4d)}\,h\mn \, ,\\
B\M&\equiv\bar g^{(d+2)/(4d)}\,C\M \, ,\\
\bar B\m&\equiv\bar g^{(d-2)/(4d)}\,\bar C\m \, .
}
\eg
Employing the new variable $f\mn$, the graviton integral
\bg
\label{eq:2.50-11}
\e^{-\Gamma_\mrm{FG}[\bar g]}=\intD_1[f_{\bigcdot\,\bigcdot}]\,\e^{-S_\mrm{FG}[h(f);\bar g]}
\eg
involves a simple measure now, but the price to be paid is the following modification of the quadratic action \gl{2.27}:
\bg
\label{eq:2.50-12}
S_\mrm{FG}[h(f);\bar g]=\foh\int\!\dd x\,f\ab\, q^{\alpha\mu}q^{\beta\nu}\widetilde{\mK}[\bar g]\mn{}\RS f\rs \, .
\eg
Concerning the transformed action \gl{2.50-12}, the following remarks are in order:\\

\noindent
\textbf{(i)} The new version of $S_\mrm{FG}$ involves a differential operator $\widetilde \mK$ which is related to the original one by a similarity transformation:
\bg
\label{eq:2.50-13}
\widetilde \mK[\bar g]\mn{}\RS=\bar g(x)^{({d-4})/({4d})}\,\mK[\bar g]\mn{}\RS\,\bar g(x)^{-({d-4})/({4d})} \, .
\eg

\noindent
\textbf{(ii)} In \Gl{2.50-12}, the volume element $\dd x$ is \ita{not} accompanied by a factor of $\sgbo$. Furthermore, as for the tensor $\bar g^{\alpha\mu}\bar g^{\beta\nu}$ that occurs under the integral of \gl{2.27}, the (inverse) background metric is replaced by a new (inverse) metric, namely
\bg
\label{eq:2.50-14}
q\MN=\bar g(x)^{+1/d}\,\bar g\MN\quad\Leftrightarrow\quad q\mn=\bar g(x)^{-1/d}\,\bar g\mn \, .
\eg
Note, however, that there is no corresponding replacement $\bar g\mn\to q\mn$ in the case of the metric dependence which enters via $\mK[\bar g]\mn{}\RS$. Hence, up to the similarity transformation \gl{2.50-13}, the operators $\mK$ and $\widetilde\mK$ have the same dependence on $\bar g\mn$.\\ \indent 
We also recall that in order to properly keep track of all $\bar g$-dependences it is crucial to always read $\mK$ (and likewise $\widetilde\mK$) as a $(2,2)$ tensor with the index structure $\mK\equiv\mK_{\bigcdot\,\bigcdot}{}^{\bigcdot\,\bigcdot}$ which allows us to interpret it as a map of $(0,2)$ tensors onto $(0,2)$ tensors.\\

\noindent
\textbf{(iii)} The definition \gl{2.50-14} entails that $\det(q_{\bigcdot\,\bigcdot})=1$, i.e., $q\mn$ is a \ita{unimodular variant of the background metric}.\\ 

\noindent
\textbf{(3) Densitized eigenmodes.} In order to evaluate \gl{2.50-11}, let us assume we have solved the eigenvalue problem \gl{eig-1} of the original kinetic operator $\mK$ and know its eigenfunctions $u_{nm}$. Then we can define new mode functions
\bg
\label{eq:2.70}
v_{nm}(x)\mn\equiv\bar g(x)^{(d-4)/(4d)}\,u_{nm}(x)\mn
\eg
and thus obtain a complete system of $\widetilde\mK$-eigenfunctions with identical eigenvalues:
\bg
\label{eq:2.71}
\widetilde\mK[\bar g]\mn{}\AB \, v_{nm}(x)\ab=\mF_n[\bar g] \, v_{nm}(x)\ab \, .
\eg
The new mode functions satisfy orthogonality and completeness relations of the form
\bg
\label{eq:2.72}
\int\dd x \,q^{\mu\alpha}(x)q^{\nu\beta}(x) \, v^*_{nm}(x)\ab\, v_{\bar n\bar m}(x)\mn=\delta_{n\bar n}\delta_{m\bar m} \, ,
\eg
\bg
\label{eq:2.73}
\sum_{n,m}q^{\alpha\rho}(x)q^{\beta\sigma}(x)\,v_{nm}(x)\rs\,v^*_{nm}(y)\mn=\delta(x-y)\delta\A_{(\mu}\delta\B_{\nu)} \, .
\eg
It is to be observed that, contrary to the simpler case where $\mK=-\Box_{\bar g}$ acts on scalars~[I], the orthogonality and completeness relations of the densitized eigenfunctions still display an \ita{explicit} dependence on the background metric, however not on the conformal factor thereof. The above relations contain no determinantal factors, and all indices are raised and lowered with the unimodular metric $q\mn$ rather than $\bar g\mn$. But, of course, the spectral problem of $\widetilde\mK[\bar g_{\bigcdot\,\bigcdot}]$, \Gl{2.71}, does ``know'' about the full background metric $\bar g\mn$, and in particular $\mF_n\equiv\mF_n[\bar g_{\bigcdot\,\bigcdot}]$ depends nontrivially also on its conformal factor, i.e., on $\sgbo$.\\ \indent
In terms of the new basis, every field configuration contributing to the functional integral \gl{2.50-11} admits an expansion
\bg
\label{eq:2.75}
f\mn(x)=\sum_{n,m}b_{nm}v_{nm}(x)\mn \, ,
\eg
and, by \gl{2.50-12}, the associated action reads in terms of the expansion coefficients:
\bg
\label{eq:2.78}
S_\mrm{FG}=\foh\sum_n\mF_n\sum_m b_{nm}^2 \, .
\eg
Furthermore, up to inessential constants which we omit, the measure turns into
\bg
\label{eq:2.79}
\mD_1[f]=\prod_{n,m}\D b_{nm} \, ,
\eg
so that it has become trivial now to evaluate the integral \gl{2.50-11}, yielding $\e^{-\Gamma_\mrm{FG}}=\prod_{n,m}\mF_n^{-1/2}$.\\

\noindent
\textbf{(4) Spectral representation of $\Gamma_\mrm{1L}$.} Thus we have shown that the one-loop graviton action equals
\bg
\label{eq:2.80}
\Gamma_\mrm{FG}[\bar g_{\bigcdot\,\bigcdot}]=\foh\sum_{n,m}\ln\lef(\mF_n[\bar g_{\bigcdot\,\bigcdot}]\ri)
\eg
wherein the $\mF_n$s are defined by \Gl{eig-1} as the spectral values of the operator $\mK[\bar g]_{\bigcdot\,\bigcdot}{}^{\bigcdot\,\bigcdot}$. Denoting those of the Faddeev-Popov operator $\mM[\bar g,\bar g]^{\bigcdot}{}_{\bigcdot}$ by $\mF_n^\mrm{gh}$, the Grassmann integral over the ghosts yields analogously
\bg
\label{eq:2.81}
\Gamma_\mrm{Fgh}[\bar g_{\bigcdot\,\bigcdot}]=-\sum_{n,m}\ln\lef(\mF_n^\mrm{gh}[\bar g_{\bigcdot\,\bigcdot}]\ri) \, .
\eg
The perfectly clearcut results \gl{2.80} and \gl{2.81} involve only objects with a well-defined interpretation, namely numbers $\mF_n[\bar g]$ and $\mF_n^\mrm{gh}$ that are determined by \gl{eig-1} and the analogous equation of the ghosts, but further potential occurences of $\bar g\mn$ are manifestly ruled out by our derivation.\\ \indent 
Moreover, this representation of the ``one-loop determinants'' makes the respective contributions of the individual eigenmodes manifest, and is therefore a suitable starting point for the construction of approximants.\\ \indent
So altogether the one-loop functional reads
\bg
\label{eq:2.82}
\Gamma_\mrm{1L}[\bar g]=\foh\sum_{n,m}\ln\lef(\mF_n[\bar g]\ri)-\sum_{n,m}\ln\lef(\mF_n^\mrm{gh}[\bar g]\ri) \, .
\eg
As we stressed in~[I] already, one better should resist the temptation to recast equations like \gl{2.82} in the common, but vague style of formal operator traces,
\bg
\label{eq:2.83}
\Gamma_\mrm{1L}[\bar g]=\foh\Tr\ln\lef(\mK[\bar g]^{\bigcdot\,\bigcdot}{}_{\bigcdot\,\bigcdot}\ri)-\Tr\ln\lef(\mM[\bar g,\bar g]^{\bigcdot}{}_{\bigcdot}\ri) \, ,
\eg
since this obscures information we are going to need later on.\\ \indent
We close this subsection with an important remark concerning Fujikawa's \ac{brst} invariant measure which we invoked for the functional integration over all metrics~\cite{grav-measure}. While the final result for the one-loop effective action, \Gl{2.82}, looks only all too familiar, it must be emphasized that this result does arise, not \ita{despite}, but rather only \ita{thanks} to the nontrivial $\bar g$-dependence of the measure. The metric dependence of the measure and of the action conspire in precisely such a way that $\Gamma_\mrm{1L}$ ends up depending on $\bar g\mn$ via the eigenvalues $\mF_n[\bar g]$ and $\mF_n^\mrm{gh}[\bar g]$ only.

\subsection{The approximants $\mathsf{App}(N)$}

In the preceeding subsetion we prepared the stage for a precise description of the approximants that we are going to consider. Invoking a $N$-cutoff now, the systems $\mathsf{App}(N)\equiv\mathsf{App}(\mathbf f(N))$ are constituted by a certain selection, yet to be specified, of $\mathbf f(N)$ modes of the graviton and ghost fields. For each value of $N=0,1,2,\cdots$, and for tensor $(\mrm{T})$ and vector $(\mrm{V})$ modes separately, this selection is described by index sets
\bg
\label{eq:2.120}
I_N^\mrm{T}\subset I^\mrm{T}\quad\text{and}\quad I_N^\mrm{V}\subset I^\mrm{V}
\eg
which give rise to truncated bases
\bg
\label{eq:2.121}
\mB_N^{\mrm{T,V}}=\lef\{u_{nm}[\bar g]\ \Big|\ (n,m)\in I_N^{\mrm{T,V}}\ri\}
\eg
that satisfy the conditions \gl{1.15}. As we insisted already, the step from $I^{\mrm{T,V}}$ to $I_N^{\mrm{T,V}}$ must not involve any metric, while the $\bar g$-dependence of the basis \ita{elements} is admissible even for an $N$-cutoff.\\ \indent
The finite quantum system $\mathsf{App}(N)$ is defined by the generating functionals \gl{2.50-2}, \gl{2.50-3}\footnote{And obvious generalizations thereof, the inclusion of sources, for instance.} whereby the functional integrals are restricted to the subspaces $\mF_N^\mrm{T,V}\equiv\mrm{span}\,\mB_N^\mrm{T,V}$.\\ \indent
A first consequence of this definition is that the one-loop effective action which describes the gravitational interaction of the system $\mathsf{App}(N)$, henceforth denoted by $\Gamma_\mrm{1L}^N\equiv\Gamma_\mrm{1L}$, is given by the following perfectly finite sum:
\bg
\label{eq:2.122}
\Gamma_\mrm{1L}^N[\bar g]=\foh\sum_{(n,m)\in I_N^\mrm{T}}\ln\lef(\mF_n[\bar g]\ri)-\sum_{(n,m)\in I_N^\mrm{V}}\ln\lef(\mF^\mrm{gh}_n[\bar g]\ri) \, .
\eg \indent
Clearly, the properties of the sequences $\{\mathsf{App}(N)\ |\ N=0,1,\cdots,\infty\}$ crucially hinge on the selection encoded in the subsets $I_N^\mrm{T,V}$. Despite the constraints \gl{1.15} there exists still a great variety of possibilities.\\ \indent
For the purposes of the present paper the following natural choice will be adopted:
\bg
\label{eq:2.123}
I_N^\mrm{T,V}=\lef\{(n,m)\in I^\mrm{T,V}\ \Big|\ n\leq N\ri\} \, .
\eg
Thus the system $\mathsf{App}(N)$ includes all field modes having eigenvalues which do not exceed $\mF_N$ or $\mF_N^\mrm{gh}$, respectively.\\ \indent
\Gl{2.123} is still a fairly conservative choice of $\mB_N^\mrm{T,V}$. With its ordering prescription ``small eigenvalues first, larger ones come later,'' it is still similar to a standard \ac{uv} cutoff. We emphasize, however, that this kind of \ita{reference to the magnitude of the eigenvalues is by no means a logical prerequisite of an $N$-cutoff.}\\ \indent
In Background Independent quantum gravity, physical ``infrared-ness'' or ``ultraviolet-ness'' are emergent, dynamically determined properties which, at best, manifest themselves in the \ita{effective} action. If the quantum effects are sufficiently strong it is not obvious therefore if they are linked to any special sort of bare field configurations \ita{under} the functional integral. In the future it should also be promising therefore to investigate sequences $\{\mathsf{App}(N)\}$ that are based upon more exotic orderings of the classical modes.\\ \indent 
In fact, recent studies along a different line of reasoning~\cite{Pagani:2019vfm,BIvac-2} seem to support this point of view. They suggest in particular relaxing the standard paradigma ``\acs{uv} (\acs{ir}) $\Leftrightarrow$ large (small) eigenvalues'' in a certain way. This is indeed something that could be done easily within our framework.

\subsection{A generalized variational identity}

For later application, let us find out how the approximants' effective action responds to a change of some arbitrary, nondynamical parameter which may be present in the classical Lagrangian of the graviton-ghost system. Clearly, the prime example of such an external ``parameter'' is the background metric, as the response of $\Gamma_\mrm{1L}[\bar g]$ to a variation $\bar g\mn\to\bar g\mn+\delta g\mn$ defines the associated stress tensor:
\bg
\label{eq:2.150}
\delta\Gamma_\mrm{1L}[\bar g]\equiv-\foh\int\dd x\,\sgbo\ T\MN[\bar g](x)\ \delta\bar g\mn(x) \, .
\eg \indent
Let us be slightly more general now and assume that ``$\delta$'' deforms the kinetic operators $\mK$ and/or $\mM$ in a certain way, and let us ask about the resulting change $\delta\Gamma_\mrm{1L}$. In our case this question is afflicted by potential subtleties which, in conventional perturbation theory, would usually be brushed over by arguing that $\Gamma_\mrm{1L}$ is of the trace-log form \gl{2.83}, and so the matrix identity
\bg
\label{eq:2.151}
\delta\Tr[\ln(\mK)]=\Tr[\mK^{-1}\delta\mK]
\eg
can be applied to it. This then would reduce the problem to the calculation of traces. However, in the present setting where the regularization of ``$\Tr$'' plays a crucial role, this reasoning is too simplistic, leaving unanswered a number of worrisome questions. (Given a certain cutoff scheme, which general properties of a trace will survive the regularization? Are they sufficient to establish a variational formula like \gl{2.151}? Are there differences between $\mathcal{P}$- and $N$-cutoffs~[I]?)\\ \indent
Our goal is to derive a representation of the variation which displays the contribution of each eigenmode of $\mK$ separately. Finite subsets of those modes constitute the degrees of freedom of the various approximants. And as we endow those systems with a quasi-physical status, we must know the individual contribution of each mode to the toal stress-energy tensor.\\ \indent
Therefore, rather than the trace-log expression, we vary the finite spectral sums in \gl{2.122} term by term:
\bg
\label{eq:2.152}
\delta\Gamma_\mrm{1L}^N[\bar g]=\foh\sum_{(n,m)\in I_N^\mrm{T}}\mF_n[\bar g]^{-1}\delta\mF_n[\bar g]-\sum_{(n,m)\in I_N^\mrm{V}}\mF_n^\mrm{gh}[\bar g]^{-1}\delta\mF_n^\mrm{gh}[\bar g] \, .
\eg
If the infinitesimal variation $\delta$ changes the kinetic operator $\mK$ to $\mK+\delta\mK$, how do its eigenvalues $\mF_n$ change then? The answer is given by the Hellmann-Feynman theorem:\footnote{For simplicity we assume that $\mF_n$ is not degenerate or, if it is, that the perturbation $\delta\mK$ does not mix different $m$-states. This will indeed be the case in our later application of the result.}
\bg
\label{eq:2.153}
\delta\mF_n[\bar g]=\bra{u_{nm}}\delta\mK[\bar g]\ket{u_{nm}} \, .
\eg
Here, the bra-ket notation refers to the $L^2$-Hilbert space structure of the field spaces, the corresponding inner product being, for symmetric $(0,2)$-tensors say,
\bg
\label{eq:2.154}
\braket{u_1}{u_2}=\int\!\dd x\,\sgbo\,\bar g\MA\bar g\NB\,u^*_1(x)\mn\,u_2(x)\ab \, .
\eg
(See Appendix A for further details.) Treating the ghost part in the same way we obtain the sought-for generalized variational identity in the following form:
\bg
\label{eq:2.155}
\delta\Gamma_\mrm{1L}^N[\bar g]=\foh\sum_{(n,m)\in I_N^\mrm{T}}\bra{u_{nm}}\mK^{-1}\delta\mK\ket{u_{nm}}-\sum_{(n,m)\in I_N^\mrm{V}}\bra{u_{nm}}\mM^{-1}\delta\mM\ket{u_{nm}} \, .
\eg \indent
In the limit $N\to\infty$, when $I_N^\mrm{T,V}\to I^\mrm{T,V}$, $\mB_N^\mrm{T,V}\to\mB^\mrm{T,V}$ so that the $u_{nm}$s form complete bases, the result \gl{2.155} obviously reproduces the standard formula $\delta\Gamma_\mrm{1L}^N=\foh\Tr[\mK^{-1}\delta\mK]-\Tr[\mM^{-1}\delta\mM]$, if the traces exist. However, the equation~\gl{2.155} makes perfect sense also for any finite number of $N<\infty$ field modes; in this case it yields the correct answer for the variation of the effective action that governs $\mathsf{App}(N)$.

\subsection{The integrated trace of the effective stress tensor}

As an application of the variational formula for \gl{2.155} that we shall need later on, let us consider the example where $\delta$ is realized by the differential operator
\bg
\label{eq:3.12}
\mT\equiv-2\int\!\dd x\,\bar g\mn(x)\frac{\delta}{\delta\bar g\mn(x)} \, .
\eg

\noindent
\textbf{(1)} To see its purpose, consider an arbitrary action-like functional of the metric, $F[\bar g]$, and associate the Euclidean stress tensor
\bg
\label{eq:3.10}
T_F\MN[\bar g]\equiv-\frac{2}{\sgbo}\frac{\delta F[\bar g]}{\delta\bar g\mn}
\eg
to it. Then, denoting the integral of its trace $\bar g\mn T_F\MN$ over the entire manifold by
\bg
\label{eq:3.11}
\Theta^F[\bar g]\equiv\int\!\dd x\,\sgbo\,\bar g\mn T_F\MN[\bar g](x) \, ,
\eg
we see that $\mT$ is nothing but the map that sends $F$ directly to the integrated trace of its stress tensor:
\bg
\label{eq:3.110}
\Theta^F[\bar g]=\mT F[\bar g] \, .
\eg 

\noindent
\textbf{(2)} A convenient way of concretely computing $\mT F$ consists in performing an infinitesimal, position-independent rescaling of the metric:
\bg
\label{eq:3.13}
\mT F[\bar g]=\frac{\D}{\D\alpha}F\lef[\e^{-2\alpha}\bar g\mn\ri]\Bigg|_{\alpha=0} \, .
\eg
This relation is easily established by Taylor expanding its \ac{rhs}.\\

\noindent
\textbf{(3)} Letting $\delta=\mT$, the variational formula \gl{2.155} yields for the application of $\mT$ to $\Gamma_\mrm{1L}^N$:
\bg
\label{eq:2.160}
\spl{
\Theta_N[\bar g]\equiv\mT\Gamma^N_\mrm{1L}[\bar g]=\ &\foh\sum_{(n,m)\in I_N^\mrm{T}}\bra{u_{nm}}\mK[\bar g]^{-1}\mT\mK[\bar g]\ket{u_{nm}}\\
&-\sum_{(n,m)\in I_N^\mrm{V}}\bra{u_{nm}}\mM[\bar g,\bar g]^{-1}\mT\mM[\bar g,\bar g]\ket{u_{nm}} \, .
}
\eg
Furthermore, explicit inspection of the kinetic operators \gl{2.24} and \gl{2.17}, respectively, reveals the scaling relations
\bg
\label{eq:2.161}
\spl{
\mK_0[\e^{-2\alpha}\bar g_{\bigcdot\,\bigcdot}]_{\bigcdot\,\bigcdot}{}^{\bigcdot\,\bigcdot}&=\e^{+2\alpha}\ \mK_0[\bar g_{\bigcdot\,\bigcdot}]_{\bigcdot\,\bigcdot}{}^{\bigcdot\,\bigcdot}\\
\ \mK_1[\e^{-2\alpha}\bar g_{\bigcdot\,\bigcdot}]_{\bigcdot\,\bigcdot}{}^{\bigcdot\,\bigcdot}&=\mK_1[\bar g_{\bigcdot\,\bigcdot}]_{\bigcdot\,\bigcdot}{}^{\bigcdot\,\bigcdot}\\
\mM[\e^{-2\alpha}\bar g_{\bigcdot\,\bigcdot},\e^{-2\alpha}\bar g_{\bigcdot\,\bigcdot}]^{\bigcdot}{}_{\bigcdot}&=\e^{+2\alpha}\ \mM[\bar g_{\bigcdot\,\bigcdot},\bar g_{\bigcdot\,\bigcdot}]^{\bigcdot}{}_{\bigcdot} \, .
}
\eg
By virtue of \Gl{3.13} they imply
\bg
\label{eq:2.162}
\spl{
\mT\mK_0[\bar g_{\bigcdot\,\bigcdot}]_{\bigcdot\,\bigcdot}{}^{\bigcdot\,\bigcdot}&=2\ \mK_0[\bar g_{\bigcdot\,\bigcdot}]_{\bigcdot\,\bigcdot}{}^{\bigcdot\,\bigcdot}\\
\mK\mK_1[\bar g_{\bigcdot\,\bigcdot}]_{\bigcdot\,\bigcdot}{}^{\bigcdot\,\bigcdot}&=0\\
\mT\mM[\bar g_{\bigcdot\,\bigcdot},\bar g_{\bigcdot\,\bigcdot}]^{\bigcdot}{}_{\bigcdot}&=2\ \mM[\bar g_{\bigcdot\,\bigcdot},\bar g_{\bigcdot\,\bigcdot}]^{\bigcdot}{}_{\bigcdot} \, .
}
\eg
In establishing the scaling behavior \gl{2.161} only the familiar identities for the Weyl transformation of the various geometric quantities are needed. It is crucial though to meticulously observe the correct positioning of any uncontracted index.\\ \indent
Thus, the variation of the full graviton kinetic operator $\mK=\mK_0+\mK_1$ is given by, in simplified notation again,
\bg
\label{eq:2.163}
\mT\mK[\bar g]=2\mK_0[\bar g]=2\mK[\bar g]-2\mK_1[\bar g] \, .
\eg
Note that $\mK_1=0$ if $\Lambda_\mrm{b}=0$, in which case $\mT\mK[\bar g]=2\mK[\bar g]$.\\

\noindent
\textbf{(4)} If we now return to \Gl{2.160}, and exploit that the $u_{nm}$s are normalized, $\braket{u_{nm}}{u_{nm}}=1$, we are led to a remarkably simple-looking and instructive result for the integrated stress tensor trace of $\mathsf{App}(N)$. Namely, 
\bg
\label{eq:2.164}
\Theta_N[\bar g]=\mT\Gamma_\mrm{1L}^N[\bar g]=\mathbf f_\mrm{G}(N)-\mathbf f_\mrm{gh}(N)-\sum_{(n,m)\in I_N^\mrm{T}}\bra{u_{nm}}\mK[\bar g]^{-1}\mK_1[\bar g]\ket{u_{nm}} \, .
\eg
Here, $\mathbf f_\mrm{G}(N)$ and $\mathbf f_\mrm{gh}(N)$ denote the approximant's total number of graviton and ghost degrees of freedom, respectively:
\bg
\label{eq:2.165}
\mathbf f_\mrm{G}(N)\equiv\sum_{(n,m)\in I_N^\mrm{T}} 1\quad,\quad
\mathbf f_\mrm{gh}(N)\equiv 2\sum_{(n,m)\in I_N^\mrm{V}} 1 \, .
\eg \indent
We observe that $\mT\Gamma_\mrm{1L}[\bar g]$ can acquire a nontrivial dependence on $\bar g\mn$ only if $\Lambda_\mrm{b}\neq 0$. Then, in \gl{2.164}, the term involving $\mK_1\propto\Lambda_\mrm{b}$ has a chance to make a nonzero contribution.\\ \indent 
For a vanishing cosmological constant, on the other hand, $\Theta_N=\mT\Gamma_\mrm{1L}$ is a perfectly metric-independent quantity. It has the remarkable property of counting the difference between the number of graviton and ghost degrees of freedom,
\bg
\label{eq:2.170}
\Theta_N=\mT\Gamma_\mrm{1L}[\bar g]=\mathbf f_\mrm{G}(N)-\mathbf f_\mrm{gh}(N)\quad\quad(\Lambda_\mrm{b}=0) \, .
\eg
In the following sections this quantity will be seen to play a central role in the backreaction of the fluctuations on the spacetime $(M,\bar g)$.

\section{Approximant systems with spherical universes}

As long as we keep the background metric fully generic, the evaluation of the energy and momentum carried by the quantum fluctuations is an extremely difficult problem that cannot be solved in closed form. For this reason we resort from now on to the same simplification we invoked in~[I] already. Namely, we restrict the allowed background geometries $(M,\bar g)$ to round $d$-spheres with an arbitrary radius, $S^d(L)$.\\ \indent 
The restriction allows us to perform all necessary calculations explicitly and exactly, and no (asymptotic) heat kernel or similar expansions are required. At the same time it leaves us with a nontrivial backreaction problem which still encapsulated the essential physics.\\ \indent
In the case $M=S^d(L)$ all metrics of interest have the form
\bg
\label{eq:3.14}
\bar g\mn(x)=L^2\gamma\mn(x)
\eg
where $\gamma\mn$ denotes the dimensionless standard metric on the \ita{unit} $d$-sphere. As a consequence, all functionals of $\bar g\mn$ turn into ordinary functions of the radius $L$ now. For any functional $F[\bar g]$ we write its restriction to spheres as $F(L)\equiv F[L^2\gamma]$. Incidentally, the operator $\mT$ acts on these functions according to
\bg
\label{eq:3.15}
\mT F(L)=-L\frac{\D}{\D L} F(L)
\eg
which is easily infered from \Gl{3.13}.

\subsection{The projected tadpole condition}

The eligible backgrounds being restricted to $S^d(L)$, the maximum symmetry of the spheres implies that the information contents of the Einstein equation \gl{3.20} is preserved when we contract it with $\bar g\MN$, and then integrate it over the entire manifold. The result reads
\bg
\label{eq:3.21}
-\frac{1}{2}(d-2)\,R(L)+d\,\Lambda_\mrm{b}=8\pi G\,\frac{\Theta_N(L)}{\mrm{Vol}[S^d(L)]}
\eg
with, respectively, the scalar curvature and volume of $S^d(L)$,
\bg
\label{eq:3.22}
R(L)=\frac{d(d-1)}{L^2}\quad,\quad\mrm{Vol}[S^d(L)]=(4\pi)^{d/2}\frac{\Gamma(d/2)}{\Gamma(d)}L^d \, .
\eg
Importantly, the integrated trace
\bg
\label{eq:3.23}
\Theta_N(L)\equiv\Theta_N[L^2\gamma]=\int_{S^d(L)}\!\dd x\,\sgbo\,\bar g\mn T\MN[\bar g](x)\Bigg|_{\bar g=L^2\gamma} 
\eg
completely specifies the energy momentum tensor in the case at hand.\\ \indent 
\Gl{3.21} is the key relation that determines the sought-for radii $L\equiv L^\text{\acs{sc}}$ of self-consistent spherical background spacetimes. Making the crucial $L$-dependences manifest it assumes the form
\bg
\label{eq:3.24}
\sigma_d\lef[d\,\Lambda_\mrm{b} L^d-\frac{1}{2}d(d-1)(d-2)L^{d-2}\ri]=8\pi G\,\Theta_N(L)
\eg
with constants $\sigma_d\equiv(4\pi)^{d/2}\Gamma(d/2)/\Gamma(d)$. In $d=4$ dimensions, for example,
\bg
\label{eq:3.25}
4\Lambda_\mrm{b}L^4-12L^2=\frac{3G}{\pi}\,\Theta_N(L) \, .
\eg\indent
The  nontrivial contents of the condition \gl{3.24} resides entirely in the function $\Theta_N(L)$, which relates to the one-loop action by $\Theta_N(L)\equiv\Theta_N[L^2\gamma]$ for $\Theta_N[\bar g]=\mT\Gamma_\mrm{1L}^N[\bar g]$. The identity \gl{3.15} shows that indeed
\bg
\label{eq:drei.10}
\Theta_N(L)=-L\frac{\D}{\D L}\Gamma_\mrm{1L}^N(L)
\eg
where
\bg
\label{eq:drei.11}
\Gamma_\mrm{1L}^N(L)\equiv\Gamma_\mrm{1L}^N[L^2\gamma]
\eg
denotes the restriction of the one-loop functional discussed in the previous section.

\subsection{An effective potential for the Hubble radius}

The restriction $\Gamma(L)\equiv\Gamma[L^2\gamma\mn]$ of the total effective action functional \gl{2.50} has the interpretation of an effective potential for the radius $L$, the Euclidean counterpart of the Hubble length. Moreover, since $S^d$ is maximally symmetric, the ``principle of symmetric criticality''~\cite{Palais} implies that the metric variation of the action functional and its restriction to $S^d(L)$ can be interchanged. Indeed, since
\bg
\label{eq:3.25}
\spl{
\Gamma(L)&\equiv\Big(S_\mrm{EH}[\bar g]+\Gamma_\mrm{1L}[\bar g]\Big)\Bigg|_{\bar g=L^2\gamma}\\
&=\frac{\sigma_d}{16\pi G}\Big\{-d(d-1)L^{d-2}+2\Lambda_\mrm{b}L^d\Big\}+\Gamma_\mrm{1L}^N(L)
}
\eg
the tadpole condition \gl{3.24} is obviously equivalent to the stationarity of this potential, i.e., to
\bg
\label{eq:3.26}
\frac{\D}{\D L}\Gamma(L)\Bigg|_{L=L^\text{\acs{sc}}}=0 \, ,
\eg
at self-consistent values of the radius, $L= L^\text{\acs{sc}}$.

\subsection{$\mathsf{App}(N)$ bases of tensor spherical harmonics}

When the background geometries are $d$-spheres, the operators $\mK[\bar g]$ and $\mM[\bar g,\bar g]$ as given in \gl{2.24} and \gl{2.17}, respectively, boil down to the Laplacian on $S^d(L)$, denoted by  $\Box_{\bar g}$, plus a constant coefficient times the unit matrix. As a consequence, $\mK[\bar g]$ and $\mM[\bar g,\bar g]$ are diagonalized by the eigenfunctions of $\Box_{\bar g}$, with eigenvalues $\mF_n$ and $\mF_n^\mrm{gh}$ given by those of $\Box_{\bar g}$, up to a constant shift.\\ \indent 
The vector and (symmetric rank-2) tensor harmonics on $S^d$ are well studied~\cite{harmonics}. In the tensor case, for instance, the eigenvalue equation
\bg
\label{eq:T.1}
-\Box_{\bar g}(u_{nm}^J)\mn=\mE_n^J\,(u_{nm}^J)\mn
\eg
admits 4 series of solutions, commonly denominated by the labels $J\in\{\mrm{TT},\mrm{TLT},\mrm{LLT},\mrm{tr}\}$. We refer to Appendix A for a brief summary of their properties and a table listing the eigenvalues $\mE_n^J$ and their multiplicities $D_n^J$.\\ 

\noindent
\textbf{(1)} For $d$-spheres the Hilbert space of square-integrable tensor fields has a basis that decomposes according to
\bg
\label{eq:T.2}
\mB_N^\mrm{T}=\mB_N^\mrm{(TT)}\cup\mB_N^\mrm{(TLT)}\cup\mB_N^\mrm{(LLT)}\cup\mB_N^\mrm{(tr)}
\eg
whereby
\bg
\label{eq:T.3}
\mB_N^{(J)}\equiv\lef\{u_{nm}^J\in\mB^{(J)}\ \Big|\ n\leq N\ ;\ m=1,\cdots,D_n^J\ri\} \, .
\eg 
For the ghosts, $\mB_N^\mrm{V}$ in \Gl{2.121} becomes the following subset of vector harmonics:
\bg
\label{eq:T.4}
\mB_N^\mrm{V}=\mB_N^\mrm{(T)}\cup\mB_N^\mrm{(L)} \, .
\eg
Here, $\mB_N^{(J)}$ is again given by \gl{T.3}, but this time for transverse $(J=\mrm{T})$ and longitudinal $(J=\mrm{L})$ vector fields.\\ \indent\newpage
Note that all graviton and ghost bases are cut off at the same highest $n$-quantum number, $n=N$, whereas the lowest possible value of $n$ depends on the type $J$. (See the table in the appendix.)\\ 

\noindent
\textbf{(2)} The system $\mathsf{App}(N)$ possesses a total of $\mathbf f_\mrm{G}(N)$ and $\mathbf f_\mrm{gh}(N)$ graviton and ghost degrees of freedom, those numbers being the cardinality of $\mB_N^\mrm{T}$ and $\mB_N^\mrm{V}$, respectively. In terms of the various multiplicities we have
\bg
\label{eq:5.15}
\mathbf f_\mrm{G}(N)=\sum_{n=2}^N\lef(D_n^\mrm{TT}+D_n^\mrm{LTT}+D_n^\mrm{LLT}\ri)+\sum_{n=1}^N D_n^\mrm{tr} \, ,
\eg
\bg
\label{eq:5.16}
\mathbf f_\mrm{gh}(N)=2\sum_{n=1}^N\lef(D_n^\mrm{T}+D_n^\mrm{L}\ri) \, .
\eg
The table in the Appendix shows that the multiplicities $D_n^J$ for different types $J$ obey the relations
\bg
\label{eq:5.166}
D_n^\mrm{LTT}=D_n^\mrm{T}\quad,\quad D_n^\mrm{LLT}=D_n^\mrm{tr}=D_n^\mrm{S}\quad\quad(n\geq 2) \, .
\eg
It is instructive therefore to express \gl{5.15} and \gl{5.16} in terms of the three independent multiplicities $D_n^\mrm{TT}$, $D_n^\mrm{T}$, and $D_n^\mrm{S}$:
\bg
\label{eq:5.17}
\mathbf f_\mrm{G}(N)=D_1^\mrm{S}+\sum_{n=2}^N\lef(D_n^\mrm{TT}+D_n^\mrm{T}+2 D_n^\mrm{S}\ri) \, ,
\eg
\bg
\label{eq:5.18}
\mathbf f_\mrm{gh}(N)=2\lef(D_1^\mrm{T}+D_1^\mrm{S}\ri)+\sum_{n=2}^N\lef(2 D_n^\mrm{T}+2 D_n^\mrm{S}\ri) \, .
\eg
\textbf{(3)} It can be observed that in calculating the difference between the numbers of graviton and ghost modes, significant cancellations occur between the terms in \gl{5.17} and \gl{5.18} on all $n$-levels with $2\leq n\leq N$. The contributions $\propto D_n^\mrm{S}$ disappear completely, for example, and we are left with
\bg
\label{eq:5.20}
\mathbf f_\mrm{G}(N)-\mathbf f_\mrm{gh}(N)=\sum_{n=2}^N\lef(D_n^\mrm{TT}-D_n^\mrm{T}\ri)-D_1^\mrm{S}-2D_1^\mrm{T} \, .
\eg
On a detailed mode-by-mode basis, this counting formula reflects how the ghosts remove the unphysical excitations from the $h\mn$ field.\\ \indent 
The cancelation mechanism is particularly transparent when $N$ is much larger than unity. Eqs. \gl{C.14} and \gl{C.15} from Appendix A yield for $N\gg 1$, in leading order,
\bg
\label{eq:5.20}
\sum_{n=2}^N D_n^\mrm{TT}=\foh(d+1)(d-2)\,\mathbf f_\mrm{scal}(N)+O(N^{d-1}) \, ,
\eg
\bg
\label{eq:5.21}
\sum_{n=2}^N D_n^\mrm{T}=(d-1)\,\mathbf f_\mrm{scal}(N)+O(N^{d-1}) \, .
\eg
Here we employ the (leading order term of the) analogous scalar sum as a convenient reference~[I]:
\bg
\label{eq:5.22}
\mathbf f_\mrm{scal}(N)\equiv\sum_n^N D_n^\mrm{S}=\frac{2}{d!}\,N^d+O(N^{d-1}) \, .
\eg
Note that the prefactors of $\mathbf f_\mrm{scal}$ in \gl{5.20} and \gl{5.21} are precisely the numbers of independent field components comprised by a transverse-traceless tensor and a transverse vector field, respectively. As a result, the difference $\mathbf f_\mrm{G}-\mathbf f_\mrm{grav}$ is seen to count exactly the number of independent polarization states of a physical graviton in $d$ dimensions:
\bg
\label{eq:5.24}
\spl{
\mathbf f_\mrm{G}(N)-\mathbf f_\mrm{gh}(N)&=\lef[\foh(d+1)(d-2)-(d-1)\ri]\mathbf f_\mrm{scal}(N)+O(N^{d-1})\\
&=\foh d(d-3)\,\mathbf f_\mrm{scal}(N)+O(N^{d-1}) \, .
}
\eg
This is indeed the correct result: a physical graviton represents $d(d-3)/2$ degrees of freedom per spacetime point\footnote{One way to see this is to note that the symmetric matrix $h\mn$ has $d(d+1)/2$ independent entries, but that only $d(d+1)/2-d-d=d(d-3)/2$ of the functions parametrizing $h\mn$ are free to independently time-evolve physical information. Thereby, the first ``$-d$'' is due to the coordinate conditions, while the second one is a consequence of Bianchi's identities which enforce unavoidable relations among the $h\mn$'s.}, or stated differently, $d(d-3)/2$ more than a real scalar would have.

\subsection{Spectral sums}

Next we evaluate the integrated trace of the stress tensor given in \Gl{2.164} for the special case of $S^d(L)$ backgrounds. Since the operator $\mK$ is diagonolized by the eigenfunctions of the tensor Laplacian on $S^d(L)$ then, and its potential-type terms are proportional to the unit operator, the eigenvalues of $\mK$ for the various series of tensor modes, $u_{nm}^J$, $J\in\{\mrm{TT},\mrm{LTT},\mrm{LLT},\mrm{tr}\}$, are readily found.\\ \indent
Restricting the calculation from now on to $d=4$ dimensions, we obtain from \gl{2.164}:
\bg
\label{eq:5.35}
\Theta_N(L)\equiv\Theta_N[L^2\gamma\mn]=\mathbf f_\mrm{G}(N)-\mathbf f_\mrm{gh}(N)+\Delta\Theta_N(L;\Lambda_\mrm{b}) \, .
\eg
This representation of the trace involves two different kinds of contributions: the (now exact) number of graviton and ghost degrees of freedom,
\bg
\label{eq:5.355}
\spl{
\mathbf f_\mrm{G}(N)&=\frac{1}{12}\lef[10 N^4+80 N^3+158 N^2-8N-180\ri] \, , \\
\mathbf f_\mrm{gh}(N)&=\frac{1}{12}\lef[8N^4+64 N^3+160 N^2+128N\ri] \, ,
}
\eg
\enlargethispage{2\baselineskip}
and a more complicated, $\Lambda_\mrm{b}$-dependent spectral sum:
\bg
\label{eq:5.36}
\spl{
\Delta\Theta_N(L;\Lambda_\mrm{b})=2\Lambda_\mrm{b}\sum_{n=2}^N\ &\Bigg\{\frac{D_n^\mrm{TT}}{\mE_n^\mrm{TT}+8L^{-2}-2\Lambda_\mrm{b}}+\frac{D_n^\mrm{LTT}}{\mE_n^\mrm{LTT}+8L^{-2}-2\Lambda_\mrm{b}}\\
&+\frac{D_n^\mrm{LLT}}{\mE_n^\mrm{LLT}+8L^{-2}-2\Lambda_\mrm{b}}\Bigg\}+2\Lambda_\mrm{b}\sum_{n=1}^N\frac{D_n^\mrm{tr}}{\mE_n^\mrm{tr}-2\Lambda_\mrm{b}} \, .
}
\eg
Several remarks are in order to this point.\\

\noindent
\textbf{(1)} The exact number counts in \gl{5.355} for $d=4$ confirm the expected asymptotics \gl{5.24},
\bg
\label{eq:5.366}
\mathbf f_\mrm{G}(N)-\mathbf f_\mrm{gh}(N)=[10-8]\frac{N^4}{12}+O(N^3) \, .
\eg
The ``10'' is due to the ten independent entries of $h\mn$, and the ``$-8$'' stems from the ghosts. This leaves us with 2 physical polarization states for the graviton, as it should be.\\ \indent
Recalling from~[I] the exact $d=4$ result for a real scalar field,
\bg
\label{eq:5.367}
\mathbf f_\mrm{scal}(N)=\frac{1}{12}\lef[N^4+8N^3+23N^2+28N\ri] \, ,
\eg
it is also interesting to note that the magic compensation ``$10-8=2$'' does not only take place in the leading $N^4$ term but also in the first subleading one which is proportional to $N^3$. In fact, the difference
\bg
\label{eq:5.368}
\lef[\mathbf f_\mrm{G}(N)-\mathbf f_\mrm{gh}(N)\ri]-2\mathbf f_\mrm{scal}(N)=O(N^2)
\eg
is of only second order in $N$ rather than third.\\ \indent 
At orders $N^k$, $k=2,1,0$, there are nonzero deviations from the ``$10-8=2$'' count. We take them as a first hint indicating that the idealization of ``quasi-physical'' approximants with finitely many degrees of freedom deteriorates for too small values of $N$. While the $N^4$ and probably also the $N^3$ terms are perfectly meaningful, lower orders should be considered with a grain of salt.\\ 

\noindent
\textbf{(2)} Note that the $\Delta\Theta_N$-contribution to $\Theta_N(L)$ is entirely due to the $h\mn$-fluctuations; the sum \gl{5.36} makes their various $J$-sectors explicit. The ghost contribution to the integrated trace consists of nothing more than the negative constant $-\mathbf f_\mrm{G}(N)$ that appears in \gl{5.35}. We had encountered this piece in Section 3 already, even without being forced there to assume any specific (i.e., simple) form of the background metric $\bar g\mn$.\\ \indent 
The contribution $\Delta\Theta_N$ with its complicated mode sum originates from the ``deformation'' of the graviton's kinetic operator $\mK=\mK_0+\mK_1$ by the term $\mK_1\propto\Lambda_\mrm{b}$. The latter introduces a second length scale into the problem that can compete with $L$ now, namely the (Hubble) length $L_\mrm{b}$ associated with the bare cosmological constant. In the case $\Lambda_\mrm{b}>0$,
\bg
\label{eq:5.37}
L_\mrm{b}\equiv\sqrt{\frac{3}{\Lambda_\mrm{b}}} \, .
\eg
\enlargethispage{2\baselineskip}
Neither $\mK_0[\bar g]$ nor the Faddeev-Popov operator $\mM[\bar g,\bar g]$ depend on this second scale.\\ \indent
If $\Lambda_\mrm{b}=0$, the integrated stress tensor for $\mathsf{App}(N)$ simplifies substantially. Since the correction term $\Delta\Theta_N$ vanishes identically in this case, the total result
\bg
\label{eq:5.40}
\Theta_N(L)=\mathbf f_\mrm{G}(N)-\mathbf f_\mrm{gh}(N)=\mrm{const}\quad\quad(\Lambda_\mrm{b}=0)
\eg
is found to be \ita{independent of the radius $L$}. Besides its sheer simplicity, also the counting properties of $\Theta_N$ are noteworthy.\\

\noindent
\textbf{(3)} Since $L$ and $\Lambda_\mrm{b}$ are the only dimensionful quantities in the game, $\Delta\Theta_N(L;\Lambda_\mrm{b})$ is actually a function of one dimensionless variable only. Henceforth assuming that $\Lambda_\mrm{b}>0$, we set
\bg
\label{eq:5.50}
\Delta\Theta_N(L;\Lambda_\mrm{b})\equiv\bm\vartheta_N(z)
\eg
where the argument of $\bm\vartheta_N$,
\bg
\label{eq:5.51}
z\equiv 2\Lambda_\mrm{b}L^2\equiv 6\lef(\frac{L}{L_\mrm{b}}\ri)^2 \, ,
\eg
measures the radius in units of the bare Hubble length.\\ \indent 
\Gl{5.36} assumes the following form now:
\bg
\label{eq:5.52}
\bm\vartheta_N(z)=z\sum_{n=2}^N\Bigg[\frac{D_n^\mrm{TT}}{n(n+3)+6-z}+\frac{D_n^\mrm{T}}{n(n+3)+2-z}
+\frac{2D_n^\mrm{S}}{n(n+3)-z}\Bigg]+\frac{5z}{4-z} \, .
\eg
Here we took advantage of the relations \gl{5.166} and inserted the eigenvalues of the tensor Laplacian. They can be found in Table~\ref{table:1}, which also provides us with the multiplicitis $D_n^J$ for general $d$. Making essential use of their explicit forms in $d=4$,
\bg
\label{eq:5.53}
\spl{
D_n^\mrm{TT}&=\frac{5}{6}(n-1)(n+4)(2n+3)\\
D_n^\mrm{T}&=\foh n(n+3)(2n+3)\\
D_n^\mrm{S}&=\frac{1}{6}(n+1)(n+2)(2n+3) \, ,
}
\eg
we can simplify \Gl{5.52} quite considerably by means of a partial fraction decomposition. It yields
\bg
\label{eq:5.54}
\spl{
\bm\vartheta_N(z)=\ &\frac{5}{3}z(N-1)(N+5)+\frac{5z}{4-z}+\frac{5}{6}z(z-10)\sum_{n=2}^N\frac{2n+3}{n(n+3)+6-z}\\
&+\foh z(z-2)\sum_{n=2}^N\frac{2n+3}{n(n+3)+2-z}
+\frac{1}{3}z(z+2)\sum_{n=2}^N\frac{2n+3}{n(n+3)-z} \, .
}
\eg \indent
Obviously, $\bm\vartheta_N(z)$ possesses three series of poles at the points $z\equiv z_{N,n}^\mrm{pole}$ where the denominators under the sums vanish:
\bg
\label{eq:4.poles}
z_{N,n}^\mrm{pole}=n(n+3)+\Delta z\quad,\quad\Delta z\in\{0,2,6\}\quad,\quad n=2,\cdots,N \, .
\eg
In Section 5 we shall come back to these poles in more detail.\\ 

\noindent
\textbf{(4)} It is instructive to reorganize \Gl{5.54} in the following style:
\bg
\label{eq:5.60}
\bm\vartheta_N(z)=\bm\vartheta_N^\mrm{quad}(z)+\bm\vartheta_N^\mrm{log}(z)+\bm\vartheta_N^\mrm{conv}(z) \, .
\eg
The three contributions to $\bm\vartheta_N(z)$ differ in their leading behavior for $N\gg 1$. If we let $N\to\infty$, at fixed $z$, the first two pieces, $\bm\vartheta_N^\mrm{quad}\propto N^2$ and $\bm\vartheta_N^\mrm{log}\propto\ln(N)$, diverge quadratically and logarithmically, respectively, while $\bm\vartheta_N^\mrm{conv}$ converges to a finite limit. The explicit expressions for $\bm\vartheta_N^\mrm{quad}$, $\bm\vartheta_N^\mrm{log}$, and $\bm\vartheta_N^\mrm{conv}$, respectively, can be found in Appendix B.\\ \indent
We emphasize in particular that \ita{$\bm\vartheta_N$ contains no quartic terms $\propto N^4$}. As a consequence, $\Delta\Theta_N(L;\Lambda_\mrm{b})\equiv\bm\vartheta_N(z)$ is suppressed by a factor of $1/N^2$ relative to the constant term in $\Theta_N$, i.e., to $\mathbf f_\mrm{G}-\mathbf f_\mrm{gh}\propto N^4$, when $N\to\infty$ at fixed $z$.\\ \indent

\noindent
\textbf{(5)} Finally, the finite sums defining $\bm\vartheta_N(z)$ can be evaluated explicitly in terms of digamma functions. We display the somewhat unwieldy result in Appendix C.

\section{Sequences of self-consistent approximants}

In this section we are going to assemble the above results in order to complete the construction of the approximant systems. The final step consists in implementing the gravitational backreaction of the quantized fluctuation degrees of freedom on the universe they live in.\\ \indent
In 4 dimensions, the Hubble radii $L\equiv L_N^\text{\acs{sc}}$ of self-consistent background universes for $\mathsf{App}(N)$ must be found from the tadpole equation
\bg
\label{eq:5.90}
4\Lambda_\mrm{b}L^4-12L^2-\frac{3G}{\pi}\Theta_N(L)=0 \, .
\eg
Thereby the integrated trace of the stress tensor,
\bg
\label{eq:5.91}
\Theta_N(L)=\mathbf f(N)+\Delta\Theta_N(L;\Lambda_\mrm{b}) \, ,
\eg
comprises a term which is leading when $N\gg 1$ and independent of $L$,
\bg
\label{eq:5.92}
\mathbf f(N)\equiv\mathbf f_\mrm{G}(N)-\mathbf f_\mrm{gh}(N)=\frac{1}{6}N^4+O(N^3) \, ,
\eg
as well as a second term which, while $L$ dependent, is subdominant with respect to $N$. In fact, for $N\to\infty$ at fixed $L$, $\Delta\Theta_N\propto N^2$ and so $\Delta\Theta_N$ is suppressed by two powers of $N$.\\ \indent
Therefore, and also because of the complexity of the function $\Delta\Theta_N$, we proceed in two steps now. At first, we derive the leading order solution to \gl{5.90} by retaining only the dominant term, i.e., the one quartic in $N$, $\Theta_N\approx\mathbf f(N)\approx\frac{1}{6}N^4$, and then we show in a second step that the essential features of the result obtained remain unaltered when we include the $\Delta\Theta_N$-correction.

\subsection{The leading-$N$ approximation}

\textbf{(1)} So, to begin with, let us neglect $\Delta\Theta_N$ and solve \Gl{5.90} with a constant $\Theta_N\equiv\mathbf f(N)$. For $\Lambda_\mrm{b}>0$, we find that the consistency condition does indeed possess regular solutions, namely exactly one self-consistent radius for each value of $N$. Expressed in terms of the bare Hubble length $L_\mrm{b}$ and the Planck length $\ell_\mrm{Pl}\equiv G^{1/2}$, the radii are given by
\bg
\label{eq:5.95}
\lef(L_N^\text{\acs{sc}}\ri)^2=\foh L_\mrm{b}^2\lef[1+\sqrt{1+\frac{1}{\pi}\lef(\frac{\ell_\mrm{Pl}}{L_\mrm{b}}\ri)^2\mathbf f(N)}\ri] \, .
\eg
The corresponding family of self-consistent backgrounds amounts to a complete sequence of approximants $\mathsf{App}(N)$, well behaved for all $N=0,1,2,\cdots,\infty$. Already its very existence is non-trivial.\footnote{See Ref.~[I] for examples with incomplete sequences or even no solutions at all.}\\ 

\noindent
\textbf{(2)} Concerning the interpretation of this sequence, recall that every approximant $\mathsf{App}(N)$ is a quantum system in its own right. It consists of $\mathbf f_\mrm{G}(N)$ quantized modes of the metric and $\mathbf f_\mrm{gh}(N)$ ghost modes, which together are effectively equivalent to $\mathbf f=\mathbf f_\mrm{G}-\mathbf f_\mrm{gh}$ modes of a \ita{physical} graviton. Those modes inhabit a background spacetime which they have selected themselves, $S^4(L_N^\text{\acs{sc}})$. Among all spheres, this is the one the modes like the most to ``live in.''\\ \indent
Symbolically, for every given $N$ the state of the toal system reads
\bg
\label{eq:5.955}
\mathsf{App}(N)\sim\Big(\mathbf f(N)\text{ physical graviton modes}\Big)\otimes S^4(L_N^\text{\acs{sc}}) \, .
\eg
Now let us move upstairs in the tower of systems and states, letting in turn $N=0$, $N=1$, $N=2,\cdots$. At the lowest level, $\mathsf{App}(0)$ is the classical system; in its universe the scale is set by the bare value of the Hubble radius, $L_0^\text{\acs{sc}}=L_\mrm{b}=\sqrt{3/\Lambda_\mrm{b}}$.\\  \indent
Then, increasing $N=1,2,3,\cdots$, \Gl{5.95} shows that the radius $L_N^\text{\acs{sc}}$ grows monotonically. Each time we add degrees of freedom to the approximant its Hubble length becomes larger.\\ \indent
In fact, if $N\gg 1$ so that $\mathbf f(N)\approx \frac{1}{6}N^4$ is a good approximation, \Gl{5.95} yields a \ita{linear} $N$-dependence of $L_N^\text{\acs{sc}}$, 
\bg
\label{eq:5.96}
L_N^\text{\acs{sc}}=\lef(\frac{1}{24\pi}\ri)^{1/4}N\sqrt{\ell_\mrm{pl}L_\mrm{b}}\lef\{1+O\lef(1\frac{1}{N}\ri)\ri\} \, .
\eg
\textbf{(3)} Most importantly, the behavior $L_N^\text{\acs{sc}}\propto N$ implies that the sequence we found possesses a well-defined limit for $N\to\infty$, and that the self-consistent backgrounds $S^4(L_N^\text{\acs{sc}})$ approach \ita{flat space} in this limit:
\bg
\label{eq:5.97}
\lim_{N\to\infty}\mathsf{App}(N)\sim(\text{fully quantized physical graviton})\otimes\lef(\text{self-consistent spacetime }\mathcal{R}^4\ri) \, .
\eg\indent
Thus our main result is that the energy and momentum of the vacuum fluctuations executed by the quantized metric, contrary to longstanding prejudices based upon background-dependent calculations, do not cause an infinite spacetime curvature. Quite the reverse is true: \ita{adding graviton degrees of freedom, i.e., lifting the cutoff, tends to flatten the universe.}\\

\noindent
\textbf{(4)} At this point let us add also some remarks on the more technical aspects of this result.\\

\noindent
\textbf{(4a)} We find that it is neither the Planck- nor the classical Hubble length that sets the scale of the $L_N^\text{\acs{sc}}$s and the spacing between them. Rather, it is the \ita{geometric mean} of those scales, $\sqrt{\ell_\mrm{b}L_\mrm{b}}$.\\

\noindent
\textbf{(4b)} Note also that upon rewriting \gl{5.96} as
\bg
\label{eq:5.99}
L_N^\text{\acs{sc}}=N\lef(\frac{G}{8\pi\Lambda_\mrm{b}}\ri)^{1/4}\lef\{1+O\lef(\frac{1}{N}\ri)\ri\} \, ,
\eg
the result is seen to be manifestly nonperturbative, displaying a non-analytic $G^{1/4}$-dependence on Newton's constant.\\

\noindent
\textbf{(4c)} The approximant system $\mathsf{App}(N)$ with its $\mathbf f(N)$ degrees of freedom is governed by the effective action
\bg
\label{eq:5.100}
\Gamma^N(L)=\frac{\pi}{3G}\lef[-6L^2+\Lambda_\mrm{b}L^4\ri]+\Gamma^N_\mrm{1L}(L)
\eg
whose one-loop contribution can be recovered from the stress tensor by integrating \gl{drei.10}:
\bg
\label{eq:5.101}
\Gamma_\mrm{1L}^N(L)=-\int^L\D L^\prime\ \frac{\Theta_N(L^\prime)}{L^\prime}+\mrm{const} \, .
\eg
In the $\Theta=\mathbf f$ approximation, this potential is strikingly simple:
\bg
\label{eq:5.102}
\Gamma^N_\mrm{1L}(L)=-\mathbf f(N)\ln(L)+\mrm{const} \, .
\eg
One easily checks that \gl{5.100} with \gl{5.102} assumes a minimum at $L=L_N^\text{\acs{sc}}$.\\ \indent
What is remarkable about \Gl{5.102} is the \ita{purely logarithmic} $L$-dependence of $\Gamma_\mrm{1L}^N$. It is the result of an exact evaluation of the one-loop determinant at order $N^4$. Specifically, $\Gamma_\mrm{1L}$ contains none of the perhaps expected terms proportional to $L^4$ and $L^2$ which, on spheres, correspond to $\int\!\!\sgo$ and $\int\!\!\sgo R$, respectively. When present, they would renormalize the values of $G$ and $\Lambda_\mrm{b}$ in the total effective action \gl{5.100}.\\ \indent 
As was discussed in~[I] already, such terms are the typical outcome of calculations which employ a dimensionful ($\mathcal P$-type) cutoff and express their answers by a (usually only asymptotic) series expansion. The chief example of a regularization scheme in this class is the heat-kernel based calculation of one-loop determinants with a proper time cutoff~[I].\\

\noindent
\textbf{(5)} In future work, it will be interesting to compare our results to those from other Background-Independent approaches such as Monte Carlo simulations of Regge calculus or CDT-based statistical models, for example~\cite{Hamber,Loll:2019rdj}. We remark however that any comparison of this kind is possible at the level of \textit{observable final results} only. The basic mathematical mechanism which we propose here is specifically related to the ``paradoxical implementation'' of Background Independence which actually does use backgrounds but fixes them dynamically. By contrast, the statistical approaches do not introduce a background at all, and so there is probably no direct technical analogue of the self-adjustment process in the form above.

\subsection{Inclusion of the $\Delta\Theta_N$ correction}

\begin{figure}[ht]
	\centering
  \includegraphics[width=0.9\textwidth]{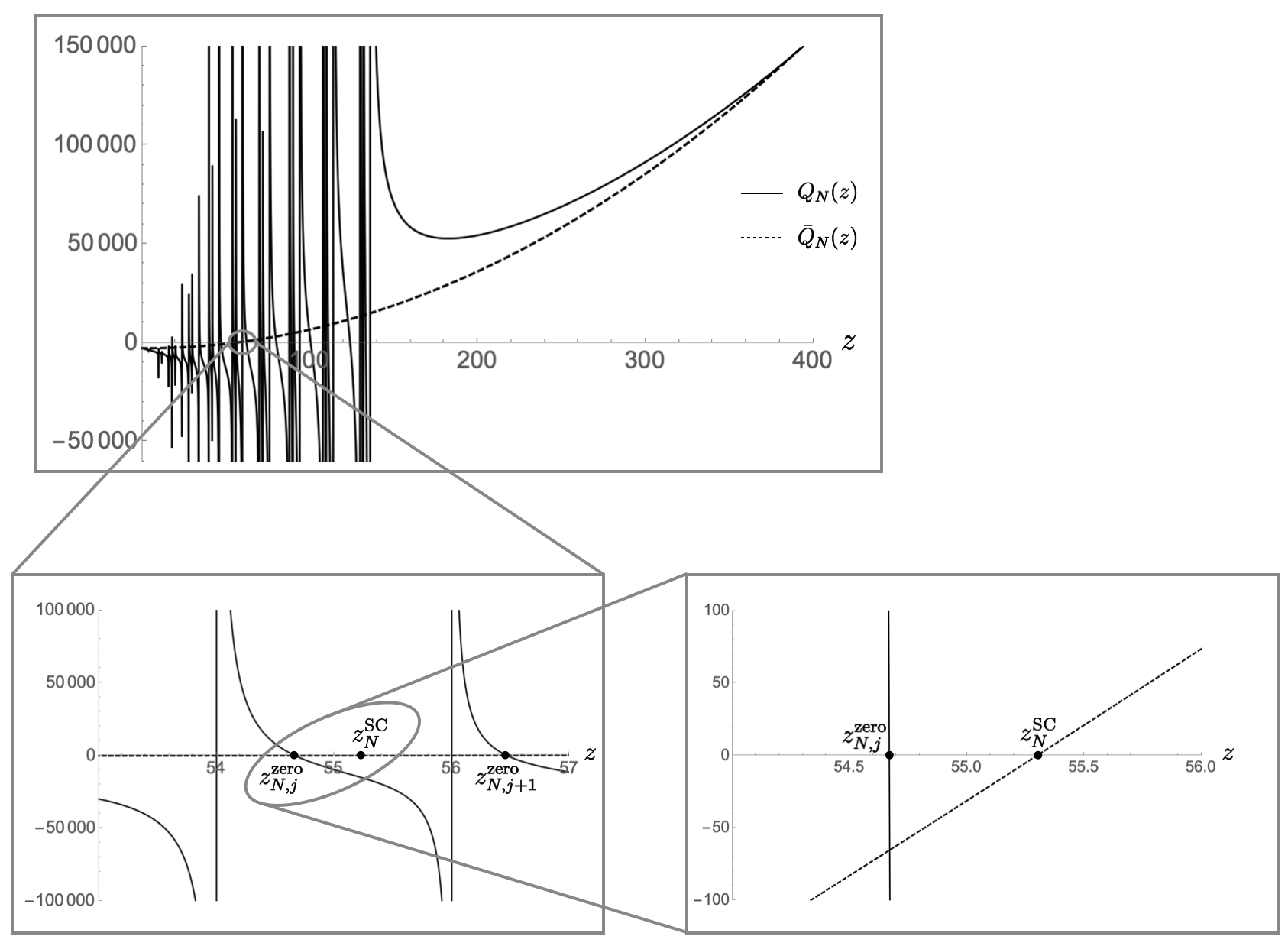}
	\caption{Graphs of the functions $Q_N(z)$ and $\bar Q_N(z)$ for $N=10$ and $G\Lambda_\mrm{b}=1$.}
	\label{fig:Q}
\end{figure}

Finally we include the correction terms $\Delta\Theta_N$, subleading with respect to $N$, into our determination of self-consistent radii. We are going to show that even in presence of $\Delta\Theta_N$ the sequence $\{\mathsf{App}(N)\ |\ N=0,1,2,\cdots\}$ from the previous subsection, with only inessential modifications, continues to be a solution to the tadpole conditions. Hence, again, the conclusion will be that spacetime approaches flat space in the continuum limit $N\to\infty$.\footnote{Interestingly, from a quite different perspective, recent work on asymptotically safe gravity~\cite{BF} found evidence which seems to point in a similar direction.}\\ \indent 
Taking advantage of the convenient variable \gl{5.51}, $z=6(L/L_\mrm{b})^2$, we write the full tadpole equation in the dimensionless form $Q_N(z)=0$, where the function $Q_N$ reads
\bg
\label{eq:5.150}
Q_N(z)\equiv z^2-6z-\frac{3}{\pi}G\Lambda_\mrm{b}\Big[\mathbf f(N)+\bm\vartheta_N(z)\Big] \, ,
\eg
and $\bm\vartheta_N(z)$ is represented by the finite sums \gl{5.54} or, in evaluated form, by \gl{5.71}. Apart from $N$, \Gl{5.150} involves only a single free parameter, namely the dimensionless product $G\Lambda_\mrm{b}$. Under the natural assumption $\Lambda_\mrm{b}=O(m_\mrm{Pl}^2)$ it is of order unity. While we shall continue to assume that $\Lambda_\mrm{b}$ is positive, our conclusions will not depend on the numerical value of $G\Lambda_\mrm{b}$.\\ \indent
With $\bm\vartheta_N$ neglected we had found only a single solution at each $N$, namely \gl{5.95}, or what is equivalent,
\bg
\label{eq:5.151}
z_N^\text{\acs{sc}}\Big|_{\bm\vartheta_N\to 0}=3\lef[1+\sqrt{1+\frac{G\Lambda_\mrm{b}}{3\pi}\mathbf f(N)}\ri]=\lef(\frac{G\Lambda_\mrm{b}}{2\pi}\ri)^{1/2}N^2+O(N) \, .
\eg
The new feature due to $\bm\vartheta_N$ is that $Q_N(z)$ possesses more than one zero now, actually a considerable number of them which increases proportional to $N^2$.\\ \indent
In Figure~\ref{fig:Q} we display the graph of $Q_N(z)$ and its analogue with $\bm\vartheta_N$ omitted, $\bar Q_N(z)\equiv z^2-6z-\frac{3}{\pi}G\Lambda_\mrm{b}\mathbf f(N)$. Obviously, the overall behavior of $Q_N$ is dictated by the polynomial $\bar Q_N$, and $Q_N$ differs from $\bar Q_N$ only close to the poles $z_{N,n}^\mrm{pole}$ located at \gl{4.poles}. There, $\bm\vartheta_N$ escapes to infinity and returns with the opposite sign, while $\bar Q_N$ is smooth. Since $\bm\vartheta_N$ is of order $N^2$ only, this feature becomes increasingly pronounced for $N\to\infty$. In this limit, $\bm\vartheta_N(z)/\mathbf f(N)\to 0$ at all regular points $z\neq z_{N,n}^\mrm{pole}$. Hence, the graph of $Q_N$ basically agrees with that of $\bar Q_N$, except for a ``forest'' of infinitely thin spikes.\\ \indent
At finite $N$, as $\bm\vartheta_N$ changes its sign at the poles, continuity implies that $\bm\vartheta_N$ must possess a zero $z_{N,j}^\mrm{zero}\in[z_{N,j}^\mrm{pole},z_{N,j+1}^\mrm{pole}]$ in between any pair of consecutive poles.\\ \indent 
In principle, any of those zeros is a possible candidate for the self-consistent radius of $\mathsf{App}(N)$. In constructing a sequence $\{\mathsf{App}(N)\ |\ N=0,\cdots,\infty\}$, we have the freedom to select, for all $N$ independently, one particular zero from the respective set of candidates: $z_N^\text{\acs{sc}}\in\{z_{N,j}^\mrm{zero}\}$.\\ \indent
\\ \indent 
The rest of the argument proceeds along the following lines now:\\

\noindent
\textbf{(1)} Different sequences of choices for $z_N^\text{\acs{sc}}$ imply different sequences of approximant systems. They may have inequivalent properties and some of them might, in principle, converge to physically different continuum limits. This situatioin is not unfamiliar: different regularization schemes may probe different (universality classes of) continuum limits.\\

\noindent
\enlargethispage{2\baselineskip}
\textbf{(2)} Near the poles the one-loop corrections are becoming huge, and this casts doubt on the reliability of the loop expansion with $\bm\vartheta_N$ included. Fortunately, the zeros are about in the middle between the neighboring poles, and so they avoid the pathological regimes as much as possible.\\

\noindent
\textbf{(3)} For specific $\mathsf{App}(N)$ sequences it is possible to diminish the influence of $\bm\vartheta_N(z)$ even further by picking the zeros for the self-consistent radii appropriately.\\ 

\indent
In fact, let us focus henceforth on the special sequence $\{\overline{\mathsf{App}}(N)\}$ which is defined by the following rule: For all $N$, select that particular zero from the set $\{z_{N,j}^\mrm{zero}\}$ which is \ita{closest to the one implied by the $\Theta_N=\mathbf f$ approximation}, i.e., to $z_N^\text{\acs{sc}}$ of \gl{5.151}. Then, denoting its label by $j=\mathbf j(N)$, the effect of $\bm\vartheta(z)$ is not more than a tiny shift by an amount $z_{N,\mathbf j(N)}^\mrm{zero}-z_N^\text{\acs{sc}}\big|_{\bm\vartheta=0}\to 0$ which decreases rapidly for $N\to\infty$ due to the proliferation of eligible zeros.\\

\noindent
\textbf{(4)} For $N\to\infty$, the entire forest of spikes and zeros, and with it the special zero $z_{N,\mathbf j(N)}^\mrm{zero}$, move towards larger values of $z$. The latter zero is very close to $z_N^\text{\acs{sc}}\big|_{\bm\vartheta=0}\propto N^2$ which runs to infinity rapidly. This, then, implies that the self-consistent spacetime related to the limit $\lim_{N\to\infty}\overline{\mathsf{App}}(N)$ has infinite Hubble radius and is nothing but flat space.\\ \indent
Thus we completed the demonstration that even if one takes the subleading $\bm\vartheta_N$ contributions at face value and includes them into the tadpole equation, there exists a ``universality class'' of $N\to\infty$ limits for which $\bm\vartheta_N$ is fully irrelevant.\footnote{Generally speaking, there could also be others. But since they would owe their existence to a nominally small correction to the leading-$N$ behavior the validity of the approximation might not extend to them.}

\section{Summary and Conclusion}

In this paper we employed a new quantization scheme in order to explore the dynamical effect which quantum vacuum fluctuations exert on the spacetimes described by metric quantum gravity. At all stages of the calculations, the scheme takes appropriate account of the pivotal role that is played by Background Independence whenever gravity is dynamical. Most importantly, the new approach extends the requirement of Background Independence to the level of the regularized precursors (``approximants'') of the quantum field theory in question and to the design of admissible cutoff schemes.\\ \indent
The general features of this approach have been outlined in Ref.~[I] already where it has also been tested for the case of classical gravity coupled to a free quantized matter field. The present investigation instead is devoted to the zero-point oscillations of the spacetime metric itself. To quantize it, we employed a one-loop approximation of quantum General Relativity, considered an effective theory. Since matter fields can be added straightforwardly, we focused on pure quantum gravity here.\\ \indent 
Our main result is that the actual effect which quantized metric fluctuations have on the curvature of spacetime is exactly opposite to what is suggested by the well-known calculations which, following Pauli~\cite{Pauli-Calc}, encounter the ``cosmological constant problem'' of a cutoff-size curvature: Each additional graviton mode that gets quantized, rather than curving spacetime more strongly, in reality \ita{reduces} the (positive, scalar) curvature, and thus drives the universee further towards flat space.\\ \indent
The explanation of this striking difference parallels the analysis in~[I], where we had obtained an analogous result for a scalar matter field. Namely, the typical standard treatments which sum zero-point energies of harmonic oscillators end end up with effective cosmological constants many orders of magnitude beyond anything acceptable, are flawed by crucially relying on the unphysical assumption of an \ita{externally provided, rigid spacetime} that would not respond to the energy and momentum of its inhabitants. As we explained in connection with Figure~\ref{fig:D} above, this assumption amounts to performing the continuum limit and including the gravitational backreaction in the wrong order. It violates Background Independence at the approximant level.\\ \indent
So, within its technical limitations,\footnote{While the explicit calculations employed the Einstein-Hilbert action, the role of Background Independence described in this paper is likely to be relevant to a much broader class of gravity theories, in particular all those admitting an effective field theory description by General Relativity at low energies. Then, at the very least, the corresponding IR fluctuation modes should exert the described effect on the cosmological constant.} the present results seem to suggest that pure Quantum Einstein Gravity should have a distinguished ground state, namely \ita{flat space}. When seen in a broader context, they make it quite obvious how dangerous it is to base investigations into the interplay of gravitational and quantum effects on the chimera of a rigid spacetime, even in cases where this violates Background Independence only at the intermediate stages of the calculation.

\vspace{0.2cm}
\noindent
\subsubsection*{Acknowledgments}
We thank Alessio Baldazzi, Kevin Falls, Renata Ferrero, Carlo Pagani, and Chris Ripken for interesting discussions. Financial support by DFG Grant RE 793/8-1 is also gratefully acknowledged.

\newpage

\begin{appendices}

\section{Tensor harmonics on $S^d$}

In this Appendix we summarize the main properties of the tensor harmonics of spin $0,1,2$ on $S^d(L)$. They satisfy the eigenvalue equation
\bg
\label{eq:C.1}
-\Box_{\bar g}\,u_{nm}=\mE_n \, u_{nm}\quad,\quad m=1,2,\cdots,D_n
\eg
with the eigenvalues $\mE_n$ and multiplicities $D_n$ listed in Table~\ref{table:1}.\\ \indent 
While there exists only one series of scalar harmonics, $u_{nm}^\mrm{S}$, in the vector case we must distinguish transverse harmonics $(u_{nm}^\mrm{T})\m$ and longitudinal ones, $(u_{nm}^\mrm{L})\m$. The former satisfy $\bar D\m(u_{nm}^\mrm{T})\M=0$, and the latter are gradients of scalar harmonics, $(u_{nm}^\mrm{L})\m=\bar D\m u_{nm}^\mrm{S}$.\\ \indent
As for symmetric rank-2 tensor harmonics, there are 4 series; three of them, namely $(u_{nm}^J)\mn$ with $J\in\{\mrm{TT},\mrm{LTT},\mrm{LLT}\}$ are comprised of traceless tensors: $\bar g\MN(u_{nm}^J)\mn=0$. The first series, the `transverse traceless' tensors $(u_{nm}^\mrm{TT})\mn$, satisfy $\bar D\M(u_{nm}^\mrm{TT})\mn=0$, while the $\mrm{LTT}$ and $\mrm{LLT}$-type tensors are derivatives of simpler harmonics:
\bg
\label{eq:C.2}
u^\mrm{LTT}{}\mn=\bar D\m u\n^\mrm{T}+\bar D\n u\m^\mrm{T} \, ,
\eg
\bg
\label{eq:C.3}
\spl{
u^\mrm{LLT}{}\mn&=\bar D\m u^\mrm{L}\n+\bar D\n u^\mrm{L}\m-\frac{2}{d}\,\bar g\mn\,\bar D\A u^\mrm{L}\a\\
&= 2\bar D\m\bar D\n u^\mrm{S}-\frac{2}{d}\,\bar g\mn\,\Box_{\bar g} u^\mrm{S} \, .
}
\eg
Finally, the tensors of the fourth series have a non-zero trace, being of the form $(u_{nm}^\mrm{tr})\mn=\bar g\mn u_{mn}^\mrm{S}$.\\ \indent
These systems of scalar, vector and (symmetric, rank-2) tensor harmonics are complete, i.e., they constitute bases of the corresponding $L^2$ Hilbert spaces. Hence a metric fluctuation $h\mn$, for example, can be expanded according to
\bg
\label{eq:C.4}
\spl{
h\mn(x)=\sum_{n=2}^\infty\Bigg\{\ &\sum_{m=1}^{D_n^\mrm{TT}}\alpha_{nm}^\mrm{TT}\,u_{nm}^\mrm{TT}(x)\mn+\sum_{m=1}^{D_n^\mrm{LTT}}\alpha_{nm}^\mrm{LTT}\,u_{nm}^\mrm{LTT}(x)\mn\\
&+\sum_{m=1}^\mrm{D_n^\mrm{LLT}}\alpha_{nm}^\mrm{LLT}\,u_{nm}^\mrm{LLT}(x)\mn\Bigg\}+\sum_{n=0}^\infty\sum_{m=1}^\mrm{D_n^\mrm{tr}}\alpha_{nm}^\mrm{tr}\,u_{nm}^\mrm{tr}(x)\mn \, .
}
\eg
For further details we must refer to~\cite{harmonics}.\\
\indent
It is instructive to consider the $n\to\infty$ asymptotics of the three independent multiplicity functions $D_n^\mrm{S}$, $D_n^\mrm{T}$, $D_n^\mrm{TT}$. In terms of the result for scalars,
\bg
\label{eq:C.10}
D_n^\mrm{S}=\frac{2}{(d-1)!}\,n^{d-1}\lef\{1+O\lef(\frac{1}{n}\ri)\ri\} \, ,
\eg
the other asymptotic degeneracies read:
\bg
\label{eq:C.11}
D_n^\mrm{T}=(d-1)\,D_n^\mrm{S}\lef\{1+O\lef(\frac{1}{n}\ri)\ri\} \, ,
\eg
\bg
\label{eq:C.12}
D_n^\mrm{TT}=\foh(d+1)(d-2)\,D_n^\mrm{S}\lef\{1+O\lef(\frac{1}{n}\ri)\ri\} \, .
\eg
The prefactors of $D_n^\mrm{S}$ in \gl{C.11} and \gl{C.12} equal precisely the numbers of independent field components which transverse vectors and transverse-traceless tensors possess in $d$ dimensions, namely $(d-1)$ and $(d+1)(d-2)/2$, respectively.
\begin{table}[ht]
\caption{Spectrum of $-\Box$ on $S^d(L)$.}
\label{table:1}
\begin{tabular}{ l l l l }
Eigenfunction & Eigenvalue & Muliplicity & $n$ \\
\hline
 & & & \\
Scalars: & & & \\
$u_{nm}$ & $\mathscr{E}_n^\mrm{S}=\frac{n(n+d-1)}{L^2}$ & $\begin{aligned}[t] {\textstyle D_n^\mrm{S}} &= {\textstyle \frac{(2n+d-1)(n+d-2)!}{n!(d-1)!}}\\ &= {\textstyle 2\binom{n+d-2}{d-1}+\binom{n+d-2}{d-2} }\end{aligned}$ & $0,1,\cdots$\\
 & & & \\
Vectors: & & & \\
 & & & \\
$\lef(u^\mrm{T}_{nm}\ri)\m$ & $\mathscr{E}_n^\mrm{T}=\frac{n(n+d-1)-1}{L^2}$ & $\begin{aligned}[t]{\textstyle D_n^\mrm{T}} &= {\textstyle \frac{n(n+d-1)(2n+d-1)(n+d-3)!}{(d-2)!(n+1)!}}\\ &= \begin{aligned}[t] &{\textstyle n\lef[\binom{n+d-1}{d-2}+\binom{n+d-2}{n}\ri]}\\ &{\textstyle +\binom{n+d-3}{n-1}}\end{aligned}\end{aligned}$ & $1,2,\cdots$\\
 & & & \\
$\lef(u^\mrm{L}_{nm}\ri)\m$ & $\mathscr{E}_n^\mrm{L}=\frac{n(n+d-1)-(d-1)}{L^2}$ & $D_n^\mrm{T}= D_n^\mrm{S}$ & $1,2,\cdots$\\
 & & & \\
Tensors: & & & \\
 & & & \\
$\lef(u^\mrm{TT}_{nm}\ri)\mn$ & $\mathscr{E}_n^\mrm{TT}=\frac{n(n+d-1)-2}{L^2}$ & $\begin{aligned}[t]
{\textstyle D_n^\mrm{TT}} 
&= {\textstyle \frac{(d+1)(d-2)(n+d)(n-1)(2n+d-1)(n+d-3)!}{2(d-1)!(n+1)!}}\\ 
&= \begin{aligned}[t] 
&{\textstyle (d+1)(n-1)\binom{n+d-3}{d-3}}\\
&{\textstyle +(d+1)(d-2)\binom{n+d-3}{d-1}}\\ 
&{\textstyle +\frac{(d+1)(n-1)}{2}\binom{n+d-2}{d-3}}
\end{aligned}
\end{aligned}$ 
& $2,3,\cdots$\\
 & & & \\
$\lef(u^\mrm{LTT}_{nm}\ri)\mn$ & $\mathscr{E}_n^\mrm{LTT}=\frac{n(n+d-1)-(d+2)}{L^2}$ & $D_n^\mrm{LTT}= D_n^\mrm{T}$ & $2,3,\cdots$\\
 & & & \\
$\lef(u^\mrm{LLT}_{n}\ri)\mn$ & $\mathscr{E}_n^\mrm{LLT}=\frac{n(n+d-1)-2d}{L^2}$ & $D_n^\mrm{LLT}= D_n^\mrm{S}$ & $2,3,\cdots$\\
 & & & \\
$\lef(u^\mrm{tr}_{n}\ri)\mn$ & $\mathscr{E}_n^{\mrm{tr}}=\mathscr{E}_n^\mrm{S}$ & $D_n^{\mrm{tr}}= D_n^\mrm{S}$ & $0,1,\cdots$
\end{tabular}
\end{table}\\ \indent
Furthermore, let us count the modes having quantum numbers $n\leq N$ for a given $N\gg 1$. The above formulae imply the following leading order behavior of the respective total number of modes:
\bg
\label{eq:C.13}
\sum_n^N D_n^\mrm{S}=\frac{2}{d!}\,N^d+O(N^{d-1}) \, ,
\eg
\bg
\label{eq:C.14}
\sum_n^N D_n^\mrm{T}=(d-1)\,\sum_n^N D_n^\mrm{S}+O(N^{d-1}) \, ,
\eg
\bg
\label{eq:C.15}
\sum_n^N D_n^\mrm{TT}=\foh(d+1)(d-2)\,\sum_n^N D_n^\mrm{S}+O(N^{d-1}) \, .
\eg
These relations are needed and discussed further in the main text.

\section{Decomposition of \texorpdfstring{$\bm\vartheta_N$}{} according to the large-$N$ asymptotics}

In this appendix we list the three terms of different large-$N$ asymptotics that appear on the \ac{rhs} of \Gl{5.60}. They are given by the following finite spectral sums:
\bg
\label{eq:5.61}
\bm\vartheta_N^\mrm{quad}(z)=\frac{5}{3}z\sum_{n=2}^N(2n+3)=\frac{5}{3}z(N-1)(N+5) \, ,
\eg
\bg
\label{eq:5.62}
\spl{
\bm\vartheta_N^\mrm{log}(z)=\ &\frac{5}{3}z(z-10)\sum_{n=2}^N\frac{n}{n(n+3)+6-z}\\
&+z(z-2)\sum_{n=2}^N\frac{n}{n(n+3)+2-z}
+\frac{2}{3}z(z+2)\sum_{n=2}^N\frac{n}{n(n+3)-z}  \, ,
}
\eg
\bg
\label{eq:5.63}
\spl{
\bm\vartheta_N^\mrm{conv}(z)=\ &\frac{5z}{4-z}+\frac{5}{2}z(z-10)\sum_{n=2}^N\frac{1}{n(n+3)+6-z}\\
&+\frac{3}{2}z(z-2)\sum_{n=2}^N\frac{1}{n(n+3)+2-z}+z(z+2)\sum_{n=2}^N\frac{1}{n(n+3)-z} \, .
}
\eg \indent
The third contribution, $\bm\vartheta_N^\mrm{conv}(z)$, approaches a well-defined limit when $N\to\infty$. It can be expressed in terms of elementary functions:
\bg
\label{eq:5.finite}
\spl{
\lim_{N\to\infty}\bm\vartheta_N^\mrm{conv}(z)=\ &\frac{\pi}{2}z\Bigg[5\frac{(z-10)}{\sqrt{4z-15}}\tan\lef(\frac{\pi}{2}\sqrt{4z-15}\ri)
+3\frac{(z-2)}{\sqrt{4z+1}}\tan\lef(\frac{\pi}{2}\sqrt{4z+1}\ri)\\
&+2\frac{(z+2)}{\sqrt{4z+9}}\tan\lef(\frac{\pi}{2}\sqrt{4z+9}\ri)\Bigg]
+\frac{5z+36}{4-z}+\frac{24}{6-z}+15z-14 \, .
}
\eg
The trigonometric functions in \gl{5.finite} give rise to infinite sequences of poles and zeros. They are simple examples of analogous sequences displayed by the general $\Psi_i$ functions discussed in Appendix C below.

\section{Representation of \texorpdfstring{$\bm\vartheta_N$}{} in terms of\\ digamma functions}

For every finite number $N$ it is possible to evaluate the spectral sums of \Gl{5.54} that represent $\bm\vartheta_N(z)$ in terms of Euler's psi-, or digamma-function. By making repeated use of the difference equation it satisfies~\cite{NIST},
\bg
\label{eq:5.70}
\psi(x+N+1)-\psi(x)=\sum_{k=0}^N\frac{1}{x+k} \, ,
\eg
we obtain the following final answer:
\bg
\label{eq:5.71}
\spl{
\bm\vartheta_N(z)=\ &\frac{5}{3}z(N-1)(N+5)+\frac{5z}{4-z}+\frac{5}{6}z(z-10)\,\Psi_1(N,z)\\
&+\frac{1}{2}z(z-2)\,\Psi_2(N,z)+\frac{1}{3}z(z+2)\,\Psi_3(N,z) \, .
}
\eg
The functions $\Psi_i$ in \gl{5.71} are given by a linear combination of digamma functions:
\bg
\label{eq:5.72}
\spl{
\Psi_i(N,z)\equiv\ &\psi\lef(N+\frac{5}{2}+\sqrt{z+\frac{1}{4}q_i}\ri)+\psi\lef(N+\frac{5}{2}-\sqrt{z+\frac{1}{4}q_i}\ri)\\
&-\psi\lef(\frac{7}{2}+\sqrt{z+\frac{1}{4}q_i}\ri)-\psi\lef(\frac{7}{2}-\sqrt{z+\frac{1}{4}q_i}\ri)  \, .
}
\eg
Herein the constants $q_i$ are $q_1=-15$, $q_2=1$, and $q_3=9$, respectively.\\ \indent 
The arguments of the digamma functions in \gl{5.72} are all real individually if $z\geq 15$, i.e.,  $L/L_\mrm{b}\geq\sqrt{5/8}$. The complete $\Psi_i$ functions are real for all $z$, however, as $\psi(x+N+1)-\psi(x)\in\mathds{R}$ for all $x\in\mathds{C}$.\\ \indent
Along the real axis, the digamma function is known to have poles of first order at the negative integers. In \gl{5.71} with \gl{5.72} they conspire so as to reproduce those of $\bm\vartheta_N(z)$ which are visible in \gl{5.54} and given in \gl{4.poles}.

\end{appendices}

\clearpage


\end{document}